\documentclass[aps,prl,notitlepage,twocolumn,superscriptaddress,longbibliography,nofootinbib]{revtex4-1}
\pdfoutput=1
\usepackage[utf8]{inputenc}
\usepackage{amsmath,amsthm,amssymb,amsfonts}
\usepackage{scalerel}
\usepackage{mathtools}
\usepackage{graphicx}
\usepackage{subfigure}
\usepackage[normalem]{ulem}
\usepackage[colorlinks = true,
            linkcolor = magenta,
            urlcolor  = magenta,
            citecolor = red,
            anchorcolor = blue]{hyperref}
\usepackage{mathrsfs}
\usepackage{bbold}
\usepackage{units}
\allowdisplaybreaks
\usepackage{bm}
\usepackage{braket}
\usepackage{makecell}
\usepackage[nameinlink]{cleveref} 
\usepackage{upgreek}
\usepackage{blindtext}
\usepackage{verbatim}
\usepackage{algorithm}
\usepackage[noend]{algpseudocode}
\makeatother

\usepackage[dvipsnames]{xcolor}
\usepackage{float}
\usepackage[bb=boondox]{mathalfa}
\usepackage{array}
\usepackage{multirow}
\usepackage{tabularx}
\usepackage{float}
\usepackage{dcolumn}
\usepackage{slashed}
\usepackage{braket}
\usepackage{verbatim}
\usepackage{multirow}
\usepackage{resizegather}
\usepackage{titlesec}
\usepackage[toc,page]{appendix}
\usepackage[export]{adjustbox}
\definecolor{teal}{RGB}{0, 158, 115} 
\definecolor{morange}{RGB}{255, 127, 0}

\DeclareMathOperator{\tr}{Tr}
\begin{document}

\title{{Information scrambling in all-to-all interacting models}}
\author{Abhik Kumar Saha\,\,\href{https://orcid.org/0000-0001-8168-6742}
{\includegraphics[scale=0.05]{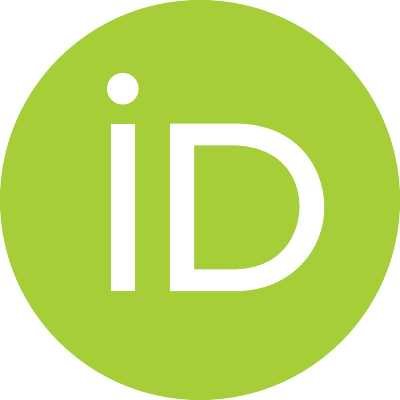}}}
\email{saha.abhikkumar.3k@kyoto-u.ac.jp}
\affiliation{Department of Physics, Kyoto University, Kitashirakawa Oiwakecho, Sakyo-ku, Kyoto 606-8502, Japan}

\author{Tanay Pathak\,\,\href{https://orcid.org/0000-0003-0419-2583}
{\includegraphics[scale=0.05]{orcidid.pdf}}}
\email{pathak.tanay.4s@kyoto-u.ac.jp}
\affiliation{Department of Physics, Kyoto University, Kitashirakawa Oiwakecho, Sakyo-ku, Kyoto 606-8502, Japan}

\author{Masaki Tezuka\,\,\href{https://orcid.org/0000-0001-7877-0839}
{\includegraphics[scale=0.05]{orcidid.pdf}}}
\email{tezuka@scphys.kyoto-u.ac.jp}
\affiliation{Department of Physics, Kyoto University, Kitashirakawa Oiwakecho, Sakyo-ku, Kyoto 606-8502, Japan}

\begin{abstract}
Information scrambling is a hallmark of quantum chaos and thermalization in isolated quantum many-body systems. We investigate scrambling dynamics in the all-to-all interacting spin Sachdev–Ye–Kitaev (SYK)-$q$ model using both pure- and mixed-state entanglement measures. We show that von-Neumann and R\'enyi entropies exhibit rapid growth followed by saturation near Haar-random values, signaling efficient scrambling. The scrambling rate reveals a nontrivial dependence on the interaction order, system size, and Hamiltonian scaling. We further employ mixed-state entanglement as a powerful probe of information scrambling. We numerically find a universal relation between the R\'enyi-1/2 mutual information and entanglement negativity for minimal interaction order in the early growth regime. Furthermore, entanglement negativity displays a Page-curve-like behavior under unequal subsystem partitioning, characterized by the birth, spread, and eventual death of quantum correlations. Our results provide a generic description of information scrambling using entanglement dynamics in all-to-all interacting spin systems with multi-body interactions.
\end{abstract}

\maketitle

\emph{Introduction.---} 
Understanding the properties of complex quantum many-body systems is a central challenge in modern physics. Specifically scrambling and related phenomena of  thermalization is important to understand the emergence of statistical mechanics in quantum many-body systems. These questions arise across many areas, from strongly correlated materials and quantum information processors to the physics of black holes. An important aspect of these studies is to choose a suitable \emph{probe} as various probes of chaos and thermalization capture different aspects of the dynamics. While spectral statistics, for example, provide powerful diagnostics of global quantum chaos, they do not directly reveal how quantum information spreads among subsystems. Entanglement offers a more microscopic perspective to understand these properties \cite{Amico:2007ag,Laflorencie:2015eck} in this regards. Pure-state measures such as von-Neumann and R\'enyi entanglement entropies characterize the growth of quantum correlations after a quench, while mixed-state probes such as mutual information, entanglement negativity \cite{PhysRevLett.77.1413,PhysRevA.65.032314,Calabrese:2012ew}, and odd entropy \cite{PhysRevLett.122.141601,Mollabashi:2020ifv,Kudler_Flam:2020url} provide access to correlations between realistic subsystems and provides finer details of how entanglement develops within the subsystem itself. These measures have therefore emerged as central tools for investigating conformal field theories \cite{PhysRevLett.109.130502,Alba_2013,Calabrese:2014yza,Chen_2017,Kudler_Flam:2020xqu,Kudler_Flam:2020url}, quantum quenches \cite{Coser:2014gsa,Alba:2018hie,Alba:2019ybw,Touil:2020vuz,Gruber:2020afu,Murciano:2021zvs,Bertini:2022fnr,Pathak:2026rfk} amongst others.

Progress on these questions often relies on theoretical models that are simple enough to analyze and at the same time rich enough to capture universal features. The Sachdev--Ye--Kitaev (SYK) has presented itself as one such model in this context. It is a model of $L$- fermions with $q$-body all-to-all random interactions \cite{PhysRevLett.70.3339,kitaev1998,kitaev2015}. It exhibits rich physics and has emerged as a quintessential model of quantum chaos, holography and beyond \cite{French:1970ztu,Bohigas:1971vpj,PhysRevX.5.041025,PhysRevB.105.075117,PhysRevB.105.235131,PhysRevB.95.155131,Gu_2020,PhysRevResearch.2.033025,PhysRevB.94.035135,PhysRevB.100.155128,Gross_2017,PhysRevD.95.026009,Li_2017,PhysRevLett.124.244101,Gates_2021,PhysRevX.12.021040,PhysRevLett.130.010401,Nandy:2024wwv,maldacena2018eternal,Jia_2022,Berkooz_2017,Gu_2017,Berkooz:2018jqr,Berkooz:2018qkz,Ozaki:PhysRevResearch.7.013092,PhysRevB.103.195108,PhysRevLett.105.151602,PhysRevLett.119.216601,Miyahara:2026iso, Maldacena:2016hyu,Garcia-Garcia:2016mno,Cotler:2016fpe,Krishnan_2018,Sarosi_2018,Maldacena:2019ufo,Trunin_2021,RevModPhys.94.035004,bousso2022snowmasswhitepaperquantum,faulkner2022snowmasswhitepaperquantum,catterall2022reportsnowmass2021theory}. Apart from its theoretical interests, it has also motivated several proposals for realizations in experimental setups \cite{Schuster_2022,Anderson_2024,Danshita_2017,Garcia-Alvarez:2016wem,Franz:2018cqi,Luo_2019,jafferis2022traversable,kobrin2025experiments,Jafferis2025,Asaduzzaman:2023wtd,Gautam:2025pnh,byun2026quantumsimulationtraversablewormholeinspiredquantum}. The models in its usual form is still \emph{complicated} and further offers simplications yet retaining essential features. In this direction spin variants of the SYK model constructed by replacing Majorana fermions with local spin operators have recently attracted considerable attention \cite{Hanada_2024,Hanada:2025pis}. Besides being more amenable to quantum simulation platforms~\cite{Gautam:2025pnh}, spin-SYK models retain several characteristic signatures of original SYK model, including spectral signatures of quantum chaos. However, replacing highly nonlocal Fermionic operators with local spin degrees of freedom raises an important question: \emph{How do interaction order and operator structure influence the dynamics of information scrambling?}

In this work, we investigate information scrambling in all-to-all interacting model taking spin-SYK-$q$ as a toy model. We characterize scrambling via entanglement production from initially unentangled state. Beyond the von-Neumann and R\'enyi entropies, we also analyze mixed-state probes such as mutual information, entanglement negativity \cite{PhysRevLett.77.1413,PhysRevA.65.032314,Calabrese:2012ew}, and odd entropy \cite{PhysRevLett.122.141601,Mollabashi:2020ifv}, which provide a more complete characterization of quantum correlations in realistic many-body subsystems. We show that the entanglement dynamics exhibits rapid growth followed by saturation near the Haar-random values at late times, signaling thermalization. The scrambling rate exhibits a nontrivial dependence on interaction order $q$, system size, and Hamiltonian scaling. Furthermore, for $q=2$, we numerically demonstrate that a universal relation between the R\'enyi-1/2 mutual information and the entanglement negativity holds throughout the early growth regime. The entanglement negativity additionally exhibit Page-curve-like behavior characterized by the birth, spreading, and eventual death of quantum correlations under subsystem partitioning. These results provide a generic description of information scrambling via entanglement spreading.

\emph{Model.---} We now introduce the spin SYK-$q$ model. 
Consider the operators \cite{Swingle:PhysRevB.109.094206,Hanada_2024,Basu:2025ubf, Pathak:2025udi}, $\hat O_{a}$, which are defined as: ${\hat O}_{2j-1}=\sigma_{j,x}$, ${\hat O}_{2j}=\sigma_{j,y}$ with $j=1,2,3..L$. Here, $\sigma_{i,k}=I_{i-1}\otimes\sigma_{i,k}\otimes I_{L-i}$ and $I_l$ denotes the $2^{l}$ dimensional identity operator. The Hamiltonian of the model is then written as  
\begin{equation}
\label{eq:spinsyskham}
H=\sqrt{\frac{(q-1)!}{(2L)^{q-1}}}\sum_{1\leq i_1<\cdots i_q\leq 2L}J_{i_1\cdots i_q}\,i^{\eta_{i_1\cdots i_q}}\prod_{k=1}^{q}\hat O_{i_k}.
\end{equation}
 where $J_{i_1...i_q}$ are standard Gaussian random variables with zero mean and unit standard deviation. $\eta_{i_1\cdots i_q}$ is the number of spins whose both $x$ and $y$ components appear in $(i_1 \cdots i_q)$ and the factor $i^{\eta_ {i_1 \cdots i_q}}$ ensures hermiticity of the Hamiltonian. Note that the Hamiltonian also has a parity symmetry for even values of $q$. In the spin SYK-$q$ model, different spin components, such as $\sigma_i^x$ and $\sigma_i^y$, appear simultaneously in the same interaction term. We can forbid this to happen and this amounts to retaining only the terms with  $\eta_{i_{1},\cdots, i_{q}}=0$ \cite{Basu:2025ubf} (see \cite{supp}).  For the numerical calculations, we consider $q= 2,3,4,5$ and $L = 16, 17$ (for odd and even $q$ respectively). Only the even parity sector is considered for even values of $q$, and full Hilbert space is taken for odd $q$. The pre-factor in the Hamiltonian, Eq.~\eqref{eq:spinsyskham}, is chosen such that the average ground state energy per site in the thermodynamic limit is a constant \cite{physprefactor} (see \cite{supp,physprefactor}).
 
\emph{Entanglement entropy.---} Consider a bi-partition of the Hilbert space $\mathcal{H}=\mathcal{H}_A\bigotimes\mathcal{H}_B$ of total dimension $\mathcal{D}$. Without loss of generality, we consider $\rm{dim}(\mathcal{H}_A)=\mathcal{N}\leq \rm{dim}(\mathcal{H}_B)=\mathcal{M}$, $\mathcal{D}=\mathcal{N}\times\mathcal{M}$. For a state $\psi_{AB} \in \mathcal{H}$, the reduced density matrix of subsystem $A$ is given as $\rho_{A}= \mathrm{Tr}_{B}(\rho_{AB})$ where $\mathrm{Tr}_{B}(\bullet)$ denotes the partial trace over subsystem $B$. $\rho_{A}$ contains the information of the amount of entanglement between subsystem $A$ and $B$, which can be obtained using $\alpha$-R\'enyi entropy ($\alpha$-RE) given as 
\begin{equation}
    S^{(\alpha)}_{A}(t)= \frac{1}{1-\alpha}\log(\mathrm{Tr}(\rho_{A}^{\alpha}))
\end{equation}
For \mbox{$\alpha\to1$}, this reduces to the usual von-Neumann entanglement entropy (EE) given as: $
S^{(\rm vN)}_{A}(t)= -\mathrm{Tr}(\rho_{A}\log(\rho_{A}))$. This is helpful to quantify the pure state entanglement. For more general case of mixed state entanglement, we will consider other measures which we discuss later. Here we only focus on the von-Neumann EE and the 2-RE. For the case of Haar random vectors the average von-Neumann \cite{Page:1993df} and the 2-RE \cite{Kim:2014Renyi} obtain a universal value given as:
$\langle S^{(\rm vN)}_{A} \rangle 
\simeq \ln(\mathcal{N}) - \frac{\mathcal{N}}{2\mathcal{M}}$ and $  
\langle S^{(2)}_{A} \rangle
\simeq \ln(\mathcal{N})
- \ln\!\left(
1 + \frac{1}{\mathcal{M}}
\left(
\mathcal{N} - \frac{1}{\mathcal{N}}
\right)
\right)
$ respectively.

To study the evolution of the von-Neumann EE and the 2-RE, we consider a product initial state \cite{PhysRevLett.111.127205} of as follows 
\begin{equation}\label{eq:pstate}
\psi_{\theta,\phi}= \bigotimes_{i=1}^{N} \left[\cos\left(\frac{\theta_{i}}{2}\right) \ket{\uparrow}+e^{i \phi_{i}}\sin\left(\frac{\theta_{i}}{2}\right) \ket{\downarrow}\right]
\end{equation}
$\theta_{i} \in [0,\pi]$ and  $\phi_{i} \in [0,2\pi]$. To remove any dependence on the initial state we always choose $\theta_{i}$ and $\phi_{i}$ to be random (see \cite{supp}). 
To study the entanglement dynamics, it is important to consider the model in a suitable timescale. There are two different timescales that one can consider: the natural timescale and the energy fixed timescale. The \emph{natural} timescale corresponds to the Hamiltonian Eq.~\eqref{eq:spinsyskham}. The energy-fixed timescale refers to the rescaling Hamiltonian, Eq.~\eqref{eq:spinsyskham}, such that the average ground state energy is equal to $-L$ \cite{gstateml}.

\begin{figure}[htbp]
	\centering
	\includegraphics[width=0.49\textwidth]{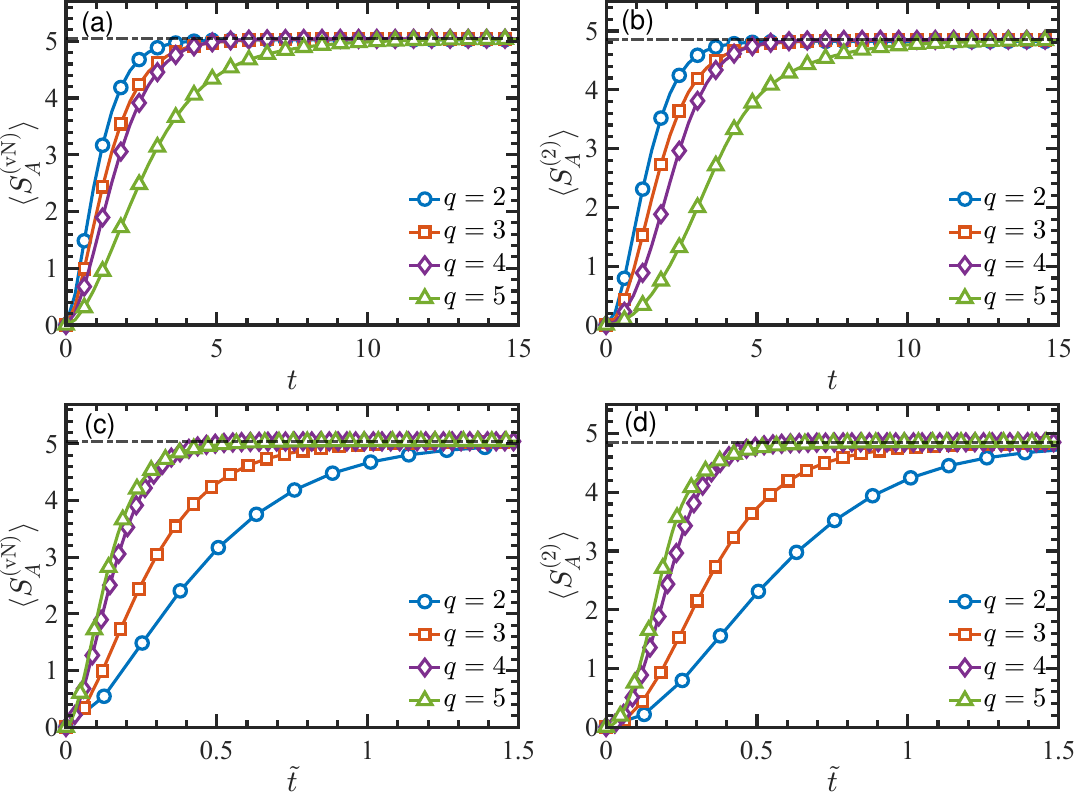}
	\caption{Evolution of (a) the average von-Neumann EE and (b) the 2-RE with time for spin-SYK model with $q=2,3,4,5$ in natural timescale. (c) and (d) shows the same in energy fixed timescale. The initial state is given by Eq.~\eqref{eq:pstate}. We consider $L=17$, even parity block for even $q$ and $L=16$ with full Hilbert space for odd $q$. The sub-systems sizes are $L_{A}=L_{B}=8$ for both the cases. The horizontal black dot--dashed line corresponds to the Haar random values. We consider $2^{20-L}$ Hamiltonian realizations and initial state is changed over each realization.
    }\label{fig:eerenyi2}
\end{figure}

In Fig.~\ref{fig:eerenyi2}, we present the results of the evolution of  the average von-Neumann EE and the 2-RE in the spin SYK-$q$ model for $q=2,3,4,5$. Here we only consider symmetric bi-partition of the system, although similar results hold for the unequal case as well (see \cite{supp}). Both entropies grow linearly from zero, reflecting the onset of many-body correlations and efficient information scrambling. A clear dependence on $q$ is observed and the entanglement growth in the natural time scale follows the ordering: $v_{2}\,\textgreater \,v_{3}\,\textgreater\, v_{4} \,\textgreater\, v_{5}$, where $v_{q}$ denotes the slope of linear growth for interaction order $q$. 
For the energy fixed time scale, on the other hand, we observe $\tilde v_2<\tilde v_3<\tilde v_4 \lesssim \tilde{v}_5$. In both cases, the dependence on $q$ is non-uniform: the successive differences between the growth rates are unequal, indicating that the scrambling rate does not vary linearly with the interaction order. At long times, both the entropies saturate to the Haar random values (dashed lines), demonstrating thermalization to a (nearly) maximally entangled state. The 2-RE saturates to a slightly lower value than the von-Neumann EE, reflecting its greater sensitivity to the larger eigenvalues of the reduced density matrix. 
 Despite quantitative differences in their saturation values, both entanglement measures display qualitatively similar dynamical behavior across all interaction orders and both time scales. Overall, the Fig.~\ref{fig:eerenyi2} demonstrates rapid entanglement growth, a clear dependence of the scrambling rate on the interaction order $q$, and saturation to the Page value indicating thermalization.   

\begin{figure}[htbp]
	\centering
	\includegraphics[width=0.49\textwidth]{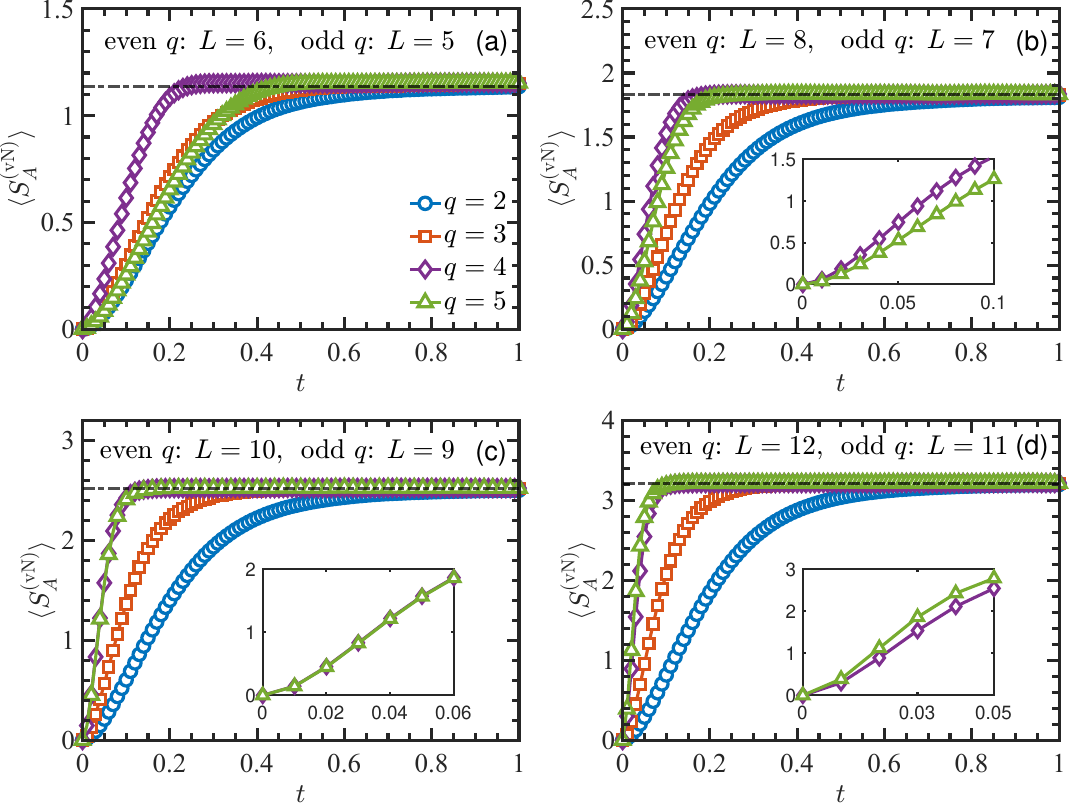}
	\caption{The dynamics of the average von-Neumann EE for different values of $q$ and (a) $L=6$ (even $q$) and $L=5$ (odd $q$), (b) $L=8$ (even $q$) and $L=7$ (odd $q$), (c) $L=10$ (even $q$) and $L=9$ (odd $q$), and (d) $L=12$ (even $q$) and $L=11$ (odd $q$), for the spin SYK-$q$ model with the prefactor set to unity. We consider $2^{20-L}$ Hamiltonian realizations. The horizontal black dot-dashed lines indicate the Haar random values. The inset shows the zoomed view of the EE in the early time regime for $q=4$ and $q=5$.}\label{fig:vnallq3}
\end{figure}

\emph{Effect of $q$ and $L$.---} Now, we analyze in detail the ordering of the entanglement growth observed previously. We first note that in a $L$ sized system with $q$-body \emph{local} interactions (not all-to-all), the growth of entanglement (quantified using linear slope) follows the ordering: $v_{2}<v_{3}< \cdots < v_{q}$; with $v_{2}$ being the slowest and $v_{q}$ being the fastest. With the inclusion of all-to-all interactions the situation becomes complicated. For this case the above ordering is not necessarily guaranteed to hold anymore as it will be a complicated combination of $q$ as well as $L$.  Hamiltonian for this case can be decomposed into local and non-local contributions as: $H_{\text{total}}= H_{\text{local}}+H_{\text{n-local}}$. In this section we show the effects of this non-local term. We study the spin-SYK-$q$ model for $q= 2, \cdots 5$ and $L = 6 (5), \cdots, 12 (11)$. We observe that that the non-local term is not enough to affect the entanglement growth for $L \sim q$ case. However, as $L$ increases the entanglement growth starts to adjust itself and finally settles curves which follow the ordering: $v_{2}<v_{3}< \cdots < v_{q}$, again. This trend is shown in Fig.~\ref{fig:vnallq3} where we take the spin SYK-$q$ with prefactor to be unity. For $q=5$ we find only at $L=12$ that we obtained the ordering $v_{2}<\cdots<v_{5}$. This ordering is robust and we also confirm this by considering a generic $p-$body spin model as well (see \cite{supp}). 

\emph{Mixed-state entanglement and Page curve.---} We next study the entanglement dynamics in the spin SYK-$q$ model in a more general setting of mixed states. Specifically, considering the tripartition of the original system as $ABC$, our interest is to study the growth of entanglement between subsystems $A$ and $B$ ($C$ is traced out). To study entanglement in such a scenario, we consider the dynamics of the following three quantities:

\emph{1.} Mutual information between two subsystems $A$ and $B$ (obtained by tracing out $C$) \cite{Witten_2020}. It is given by
\begin{equation}
    I^{(\alpha)}_{A:B}(t)= S_{A}^{(\alpha)}(t)+ S_{B}^{(\alpha)}(t)-S_{AB}^{(\alpha)}(t)
\end{equation}
where $S^{(\alpha)}_{A}(t)$ denotes the $\alpha$-R\'enyi entropy.

\emph{2.} Entanglement negativity \cite{PhysRevLett.77.1413,Calabrese:2012ew} which is defined as 
\begin{equation}
    \mathcal{E}(t)= \ln \tr\left(\sqrt{(\rho_{AB}^{T_{B}}(t))^{\dagger}\rho_{AB}^{T_{B}}(t)}\right)
\end{equation}
where $\rho_{X\Bar{X}}^{T_{\bar{X}}}$ denotes the partial transposition of subsystem $\bar{X}$.

\emph{3.} Odd entropy \cite{PhysRevLett.122.141601,Mollabashi:2020ifv,Kudler_Flam:2020url} also uses the information of the sign of the eigenvalues of $\rho_{AB}^{T_{B}}$ and is defined as 
\begin{equation}
    \mathcal{E}^{(o)}(t)= -\sum_{\lambda_{i}>0}|\lambda_{i}|\log(|\lambda_{i}|)+ \sum_{\lambda_{i}<0}|\lambda_{i}|\log(|\lambda_{i}|),
\end{equation}
where $\lambda_{i}$ are the eigenvalues $\rho_{AB}^{T_{B}}(t)$. 

\begin{figure}[htbp]
	\centering
	\includegraphics[width=0.49\textwidth]{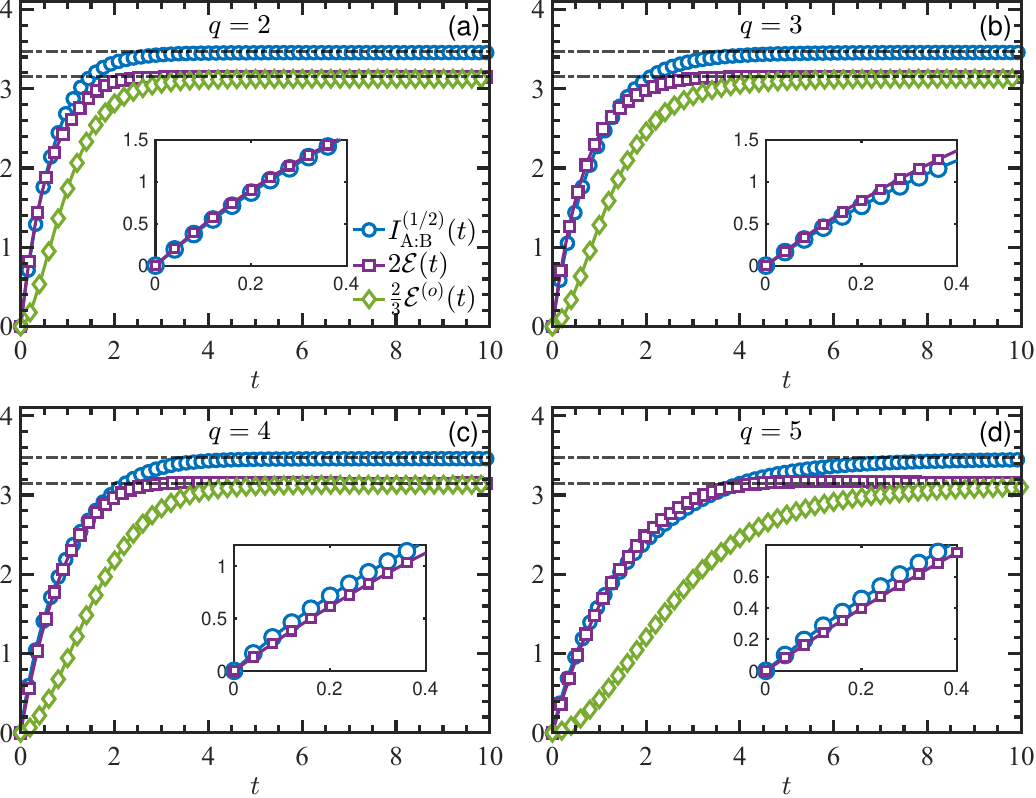}
	\caption{The dynamics of the average of $(I_{A:B}^{(1/2)}(t))$, $(2\mathcal{E}(t))$ and $(\frac{2}{3}\mathcal{E}^{(o)}(t))$ for the spin SYK-$q$ model, with (a) $q=2$, (b) $q=3$, (c) $q=4$, (d) $q=5$ and $2^{20-L}$ Hamiltonian realizations each. For even $q$ ((a) and (c)), we choose $L=16$ with even parity block and equal subsystem sizes $L_A=L_B=L_C=5$. For odd $q$ ((b) and (d)), we consider $L=15$ (full Hilbert space) with subsystem sizes  $L_A=L_B=L_C=5$. The horizontal black dot-dashed lines indicate to the Haar random values. The inset shows the zoomed view of in the growth regime for different values of $q$.}\label{fig:mutualnegodd}
\end{figure}

\begin{figure}[htbp]
	\centering
	\includegraphics[width=0.49\textwidth]{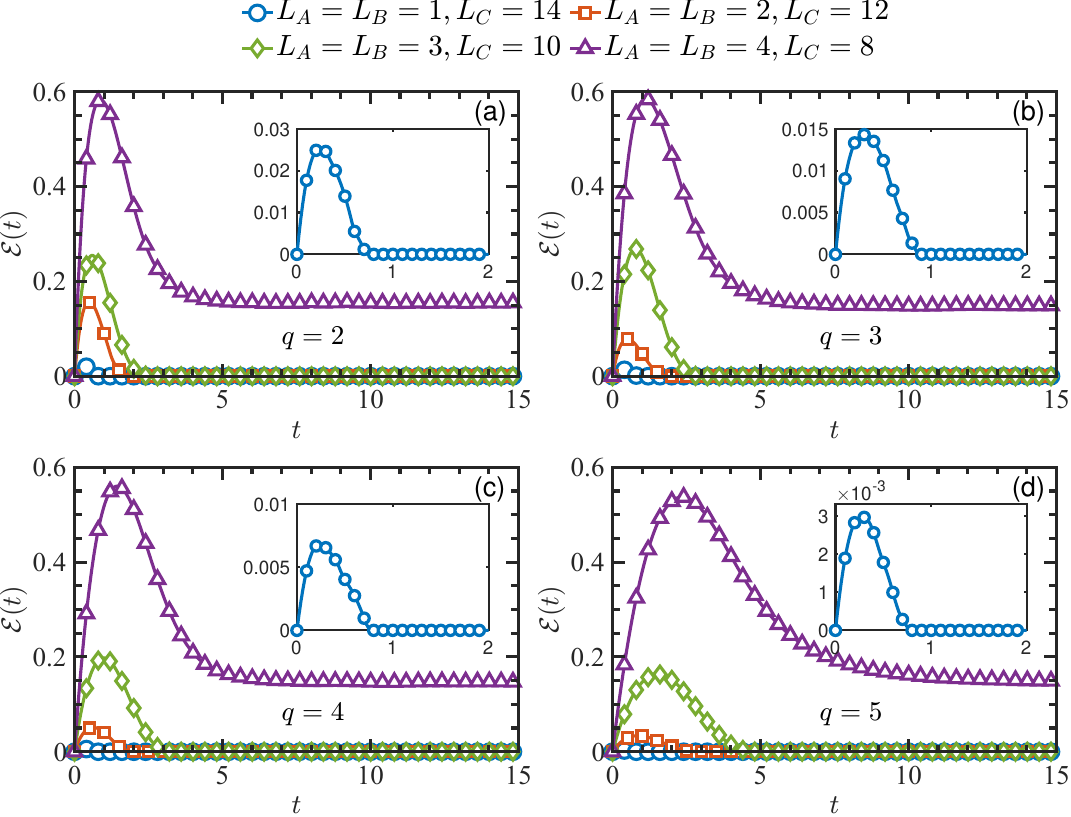}
	\caption{The dynamics of the average $(\mathcal{E}(t))$ for spin SYK-$q$ model with (a) $q=2$, (b) $q=3$, (c) $q=4$, (d) $q=5$ and initial state is given by Eq.~\eqref{eq:pstate}. Different colors and markers represent different subsystem sizes. The inset shows the zoomed view of $\mathcal{E}(t)$ for $L_A=L_{B}=1, L_C=14$ in the early time regime for different values of $q$. We consider $2^{20-L}$ Hamiltonian realizations.}\label{fig:negpagecurve}
\end{figure}

In Fig.~\ref{fig:mutualnegodd} we show the behavior of the averaged $I_{A:B}^{(1/2)}(t)$, $2\mathcal{E}(t)$ and $\frac{2}{3}\mathcal{E}^{(o)}(t)$ (see \cite{negoddrescale}) for different values of $q$ (we omit $\braket{\cdot}$ for readability). All the entanglement measures show an earlier growth regime followed by a saturation to the Haar values at late times. Another notable feature is that for the case of $q=2$ we numerically found the relation 
$$I_{A:B}^{(1/2)}(t)= 2 \mathcal{E}(t)$$ to hold during the growth regime, as shown in the inset. Such a relation has been previously identified in conformal field theories (CFTs) \cite{Kudler_Flam:2020xqu,Kudler_Flam:2020url,Murciano:2021zvs} and certain many-body models \cite{Alba:2018hie,Gruber:2020afu,Murciano:2021zvs,Bertini:2022fnr,Pathak:2026rfk}. The relation is observed partially for $q=3$ and starts to show systematic deviations for $q=4,5$. The relation is expected to hold in general for local interacting systems \cite{Bertini:2022fnr} such that the two edges of the subsystems, at early times are causally disconnected. This then implies that for $q=2$ there is \emph{finite} speed of propagation of information. The odd-entropy and the entanglement negativity agree at late times, equaling to the Haar random values. Furthermore, all the mixed-state entanglement measures, in natural timescale are found to follow the ordering: $v_2>v_3\gtrsim v_4>v_5$. For the energy fixed time scale the ordering is: $\tilde v_2< \tilde v_3<\tilde v_4\lesssim \tilde v_5$. The ordering is observed after the initial transient regime. One can also consider the case where $L_{C}\gg L_{A},L_{B}$. As shown in Fig. \ref{fig:negpagecurve}, we observe that for this case the negativity increases as time progresses, reaches a maximal value and eventually goes to zero at late times. This indicates that for this situation subsystem $AB$ factorizes. We also observe that for the case when $L_{A},L_{B}= 4, L_{C}= 8$ the late time value is non-zero which we attribute to the finite-size effects. This behavior is reminiscent of the Page curve, of relevance in black hole physics \cite{Page:1993wv,Page:2013dx}.

\emph{Outlook and discussion.---} 
In this work, we have studied in detail the pure and mixed state entanglement dynamics of SYK-$q$ model for varying $q$. 
We compare the entanglement production rates for two different time scales: natural and energy fixed. While for the natural timescale the $q=2$ scrambles the fastest, the trend is reversed for the case of energy fixed timescale. The reason is that for the natural time scale the Hamiltonian has a $q$ dependent prefactor that changes the properties of model across $q$ while $L$ is fixed. For the case of energy fixed timescale however this non-trivial dependence on $q$ is removed. It is further revealed that due to the all-to-all nature of the model an optimal value of $L$ is required to see the proper effects of interaction order $q$. Furthermore, we confirm these properties for the case of mixed state entanglement production as well. Along the way we also numerically observe that for $q=2$ previously known relation: $I_{A:B}^{(1/2)}(t)= 2\mathcal{E}(t)$, is satisfied for the growth regime. The relation is partially valid for $q=3$ as well and start to show visible deviations for $q=4$, and $5$. Furthermore for the case when a subsystem $C$ is very large as compared to other two subsystems, $A$ and $B$, it is found that the entanglement negativity increases rapidly, reaches a maximal value, followed by a decay and eventually reaches zero at late times. This behavior is reminiscent of the Page curve behavior of interest in black hole physics. It implies that the subsystem $AB$ is factorizable at late times. We argue that these features are generic and robust and should be applicable to other all-to-all interacting models in general. As a future direction it would be important to further consider the effect of the local Hilbert space dimensions, by considering SU$(d)$ matrices, on the entanglement production rates. $d=3$ is especially important as it is observed that for the case of a one-dimensional chain having all-to-all interactions, the best performance in quantum computers is shown by qutrits, with local dimension $d=3$ \cite{PhysRevA.101.022304}.

\emph{Acknowledgments.---} T.~P. and M.~T. gratefully acknowledge support from JST CREST (Grant No. JPMJCR24I2).
A.~K.~S. and M.~T. gratefully acknowledge support from JSPS KAKENHI Grant Number JP25K00925. Numerical computations were partially performed using the computational facilities of the Yukawa Institute for Theoretical Physics.

\twocolumngrid
\bibliography{references}

\begin{thebibliography}{103}
\bibitem [{phy()}]{physprefactor}%
  \BibitemOpen
  \href@noop {} {}\bibinfo {note} {Note that this is a physical condition in the sense that the total energy of the subsystem is an extensive quantity. Other conditions which ensure that the energy does not scale with system size are considered unphysical.}\BibitemShut {Stop}%
\bibitem [{gst()}]{gstateml}%
  \BibitemOpen
  \href@noop {} {}\bibinfo {note} {This can be numerically achieved by using the fact that for ground state energy, $E_{0}$, $\frac{\vert E_{0}\vert}{L}= C(q)$. This is attained in the $L\to\infty$ limit. where $C(q)$ is a $q$ dependent constant. We can then rescale our Hamiltonian given by Eq. \eqref{eq:spinsyskham} by $C(q)$.}\BibitemShut {Stop}%
\end{thebibliography}
\end{document}


\title{{Supplemental Materials: Information scrambling in all-to-all interacting models}}
\author{Abhik Kumar Saha\,\,\href{https://orcid.org/0000-0001-8168-6742}
{\includegraphics[scale=0.05]{orcidid.pdf}}}
\email{saha.abhikkumar.3k@kyoto-u.ac.jp}
\affiliation{Department of Physics, Kyoto University, Kitashirakawa Oiwakecho, Sakyo-ku, Kyoto 606-8502, Japan}

\author{Tanay Pathak\,\,\href{https://orcid.org/0000-0003-0419-2583}
{\includegraphics[scale=0.05]{orcidid.pdf}}}
\email{pathak.tanay.4s@kyoto-u.ac.jp}
\affiliation{Department of Physics, Kyoto University, Kitashirakawa Oiwakecho, Sakyo-ku, Kyoto 606-8502, Japan}

\author{Masaki Tezuka\,\,\href{https://orcid.org/0000-0001-7877-0839}
{\includegraphics[scale=0.05]{orcidid.pdf}}}
\email{tezuka@scphys.kyoto-u.ac.jp}
\affiliation{Department of Physics, Kyoto University, Kitashirakawa Oiwakecho, Sakyo-ku, Kyoto 606-8502, Japan}

\maketitle
In this supplemental material, we provide the additional results supporting the claims of the main text. In particular:
\begin{itemize}
\item In Section \ref{ssec:initialstate} we show initial state dependence for the case of spin SYK-$4$ model. 
\item In Section \ref{ssec:initialstateq4} we show initial state dependence for the usual SYK model.
    \item In Section \ref{ssec:eetgenstate} we show additional results for the pure state entanglement measures.
    \item In Section \ref{ssec:mixedstate} we show additional results for the mixed state entanglement measures.
    \item In Section \ref{ssec:pspin} we discuss the properties of the $p$- spin model.
    \item In Section \ref{ssec:gsyk} we discuss the pure state entanglement measures for genuine spin SYK-$q$ model.
    \item In Section \ref{ssec:dos} we provide the density of state of the spin SYK-$q$ model for various $q$ along with other numerical details.
\end{itemize}

\section{Initial state dependence in Spin SYK-4 model}\label{ssec:initialstate}
In this section we provide the results for the dependence of initial state for the spin SYK-$q$ model. The Hamiltonian of the model is given by 
\begin{equation}\label{seq:spinsyskham}
 H=\sqrt{\frac{(q-1)!}{(2L)^{q-1}}}\sum_{1\leq i_1<\cdots i_q\leq 2L}J_{i_1\cdots i_q}\,i^{\eta_{i_1\cdots i_q}}\prod_{k=1}^{q}\hat O_{i_k}.
\end{equation}

\begin{figure}[htbp]
	\centering
	\includegraphics[width=7cm]{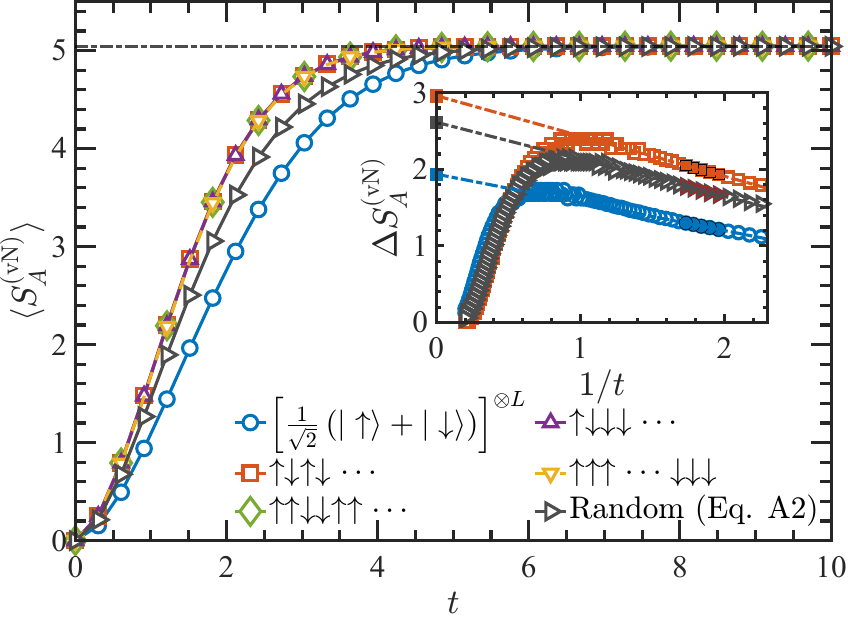}
	\caption{The dynamics of the average von-Neumann EE for the spin SYK model with $q=4$ for different initial states and averaged over $2^{20-L}$ Hamiltonian realizations. The system size considered here is $L=17$ in the even-parity sector, with equal subsystem sizes $L_{A}=L_{B}=8$,. The horizontal black dot-dashed line indicates the corresponding Haar random values given. The inset shows the instantaneous slope $\Delta S_{A}^{\rm (vN)}(t)$, plotted as a function of $1/t$. Filled markers indicate the approximate linear regime used for fitting, while the dot-dashed lines show linear extrapolations based on five selected points. The extrapolated lines intercept $1/t= 0$ and provide the asymptotic entropy production rate, highlighting the dependence of scrambling dynamics on the choice of initial states.}\label{sfig:ee_initial_state}
\end{figure}

We choose the initial states of the following form \cite{PhysRevLett.111.127205} as follows: 
\begin{equation}\label{seq:pstate}
\psi_{\theta,\phi}= \bigotimes_{i=1}^{N} \left[\cos\left(\frac{\theta_{i}}{2}\right) \ket{\uparrow}+e^{i \phi_{i}}\sin\left(\frac{\theta_{i}}{2}\right) \ket{\downarrow}\right]
\end{equation}
$\theta_{i} \in [0,\pi]$ and  $\phi_{i} \in [0,2\pi]$.

We will be interested in average von-Neumann EE ($\langle S^{(\rm vN)}_{A} \rangle$) and $2$-RE ($\langle S^{(2)}_{A} \rangle$). The Haar random values of the von-Neumann EE and $2$- RE defined as
\begin{align}
\langle S^{(\rm vN)}_{A} \rangle 
&\simeq \ln(\mathcal{N}) - \frac{1}{2Q}, 
\qquad 
Q = \frac{\mathcal{M}}{\mathcal{N}},
\label{seq:von-analytical}  \\
\langle S^{(2)}_{A} \rangle
&\simeq \ln(\mathcal{N})
- \ln\!\left(
1 + \frac{1}{\mathcal{M}}
\left(
\mathcal{N} - \frac{1}{\mathcal{N}}
\right)
\right).
\label{seq:renyi-analytical} 
\end{align}

Fig.~\ref{sfig:ee_initial_state} demonstrates the dynamics of the $\langle S_{A}^{\rm (vN)}\rangle$ for the spin SYK-$4$ model for several experimentally motivated initial states. Specifically we choose the following state-  
\begin{enumerate}
    \item Random state : $\psi_{R}=\bigotimes_{i=1}^{N} \left[\cos\left(\frac{\theta_{i}}{2}\right) \ket{\uparrow}+e^{i \phi_{i}}\sin\left(\frac{\theta_{i}}{2}\right) \ket{\downarrow}\right]$, $\theta_{i} \in[0,\pi] $ and $\phi_{i} \in[0,2\pi]$ are chosen randomly. 
    \item Symmetric superposition state : $\psi_{S}= \bigotimes_{n=1}^{N} \left(\frac{\ket{\uparrow} + \ket{\downarrow}}{\sqrt{2}}\right)$; corresponding to $\theta_i=\pi/2$ and $\phi_i=0$.
    \item N\'eel state:  $|\psi_1\rangle = |\uparrow \downarrow \uparrow \downarrow \cdots \rangle$, $\theta_{i}=0$ for odd sites and $\theta_{i}=\pi$ for even sites, with $\phi_{i}=0$ for all sites.
    \item The double N\'eel state: $|\psi_2\rangle = |\uparrow \uparrow \downarrow \downarrow \uparrow \uparrow\cdots \rangle$, $\theta_i=0$ for $i=4n+1,4n+2$ and $\theta_i=\pi$ for $i=4n+3,4n+4$, with $\phi_i=0$ for all sites.
    \item The single excitation state: $|\psi_3\rangle = |\uparrow \downarrow \downarrow \downarrow \downarrow \cdots \rangle$, $\theta_1=0$ and $\theta_i=\pi$ for $i=2,3,\ldots,L$, with $\phi_i=0$ for all sites.
    \item The domain-wall state : $|\psi_4\rangle = |\uparrow \uparrow \uparrow \cdots \downarrow \downarrow  \downarrow \rangle$, corresponding to $\theta_i=0$ for $i=1,2,\ldots,L/2$, $\theta_i=\pi$ for $i=L/2+1,\ldots,L$, and $\phi_i=0$ for all sites.
\end{enumerate}

For all the initial states, the von-Neumann EE exhibits a linear initial growth followed by a gradual approach towards the Haar random value at late times. However, the superposition state, random state, and the remaining four states exhibit distinct growth rates during the early-time regime. Notably, four out of six, initial states nearly overlap with each other throughout the dynamics, yet they remain clearly separated from the other two, namely the superposition and the random initial states. The symmetric superposition initial state exhibits the slowest growth of von-Neumann EE, indicating that the coherent superposition structure delays the scrambling of quantum information. Physically, the strong interference between the symmetry-related components of the wavefunction suppresses the rapid generation of many-body correlations, leading to slower entanglement production. In contrast, the experimentally motivated product states thermalize more efficiently and therefore show comparatively faster entanglement growth. The random initial state lies between these two extremes: indicating the behavior of a \emph{typical} state from the ensemble of pure states. 

We further conduct a more refined analysis of the entanglement growth rate by analyzing the instantaneous slope defined as:  
$$\Delta S_{A}^{\rm (vN)}((t_{i-1}+t_{i})/2)=(S_{A}^{\rm (vN)}(t_{i})-S_{A}^{\rm (vN)}(t_{i-1}))/(t_{i}-t_{i-1})$$ 
as shown in the inset of Fig.~\ref{sfig:ee_initial_state}. We observe the linear dependence of instantaneous slope on $1/t$ for all initial states. This behavior suggests the presence of logarithmic corrections to the entanglement growth. Importantly, the refined analysis of slope shows strong dependence on the initial state. Consequently, different experimentally relevant initial states display distinct entanglement-growth characteristics despite the all-to-all interacting nature of the Hamiltonian.
\clearpage

\section{Initial state dependence in usual SYK model}\label{ssec:initialstateq4}
\begin{figure}[h]
	\centering
	\includegraphics[width=7cm]{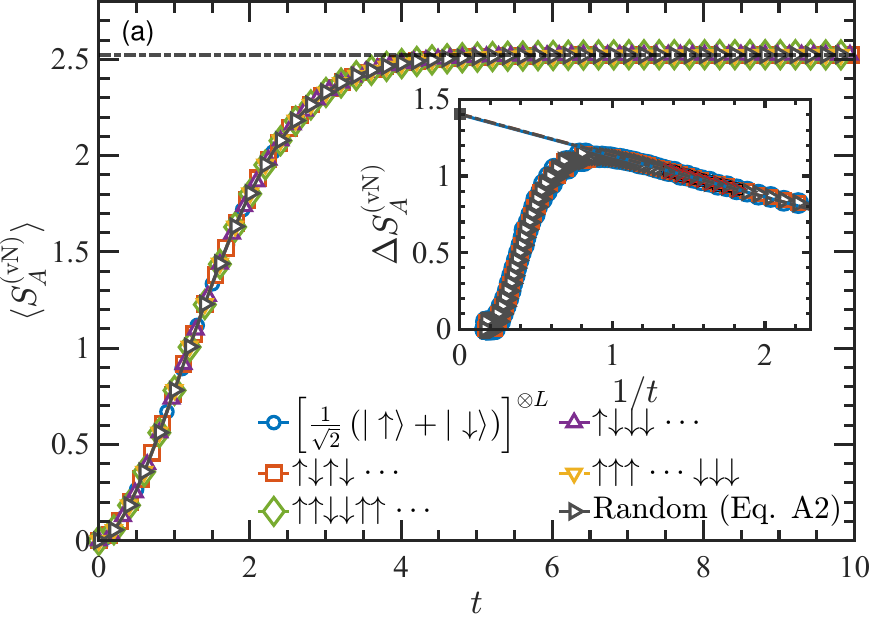}
    \includegraphics[width=7cm]{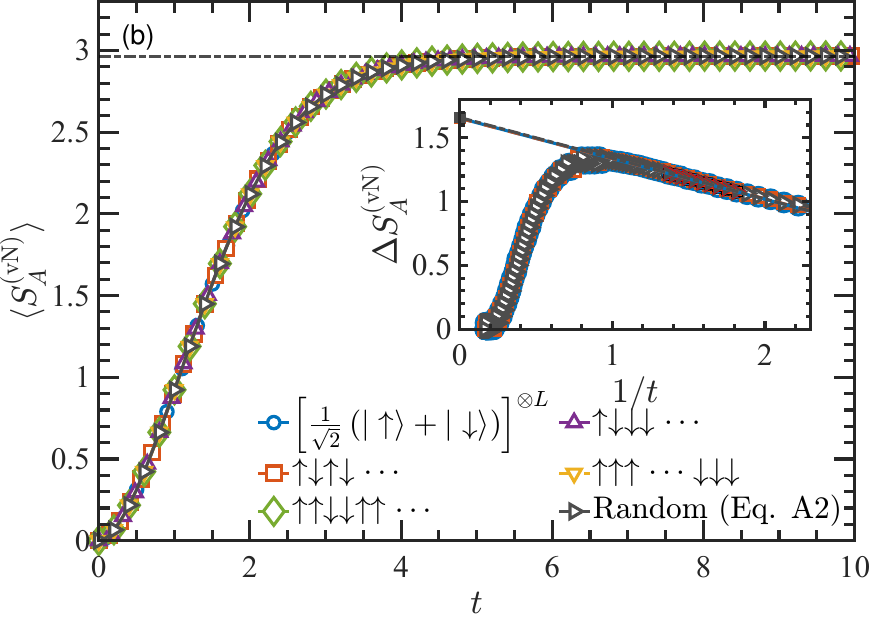}
    \includegraphics[width=7cm]{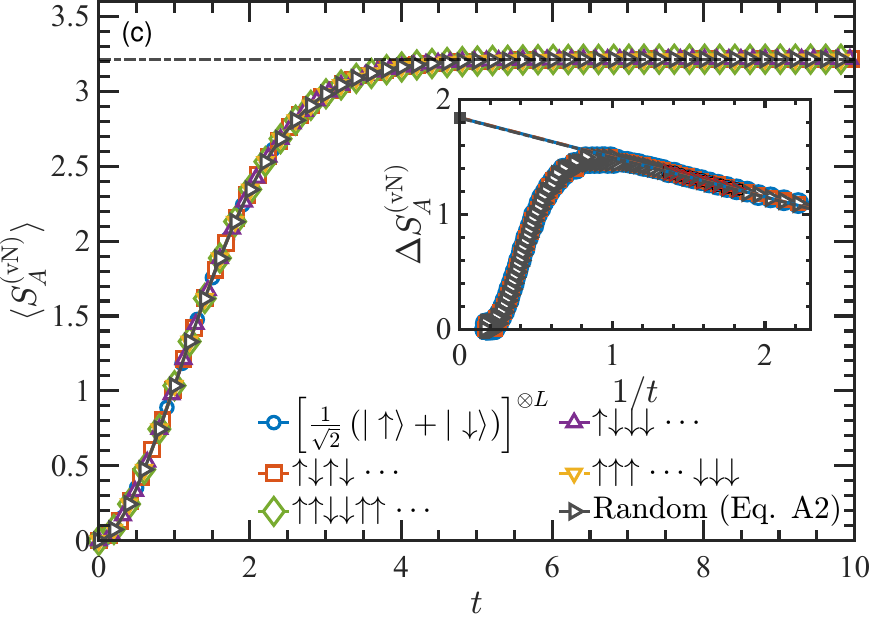}
	\caption{The dynamics of average von-Neumann EE for the usual SYK model with $q=4$ for different initial states are shown for system sizes: (a) $N=2L=20$ in the even-parity sector with subsystem sizes $L_{A}=4$ and $L_{B}=5$, (b) $N=2L=22$ in the even-parity sector with subsystem sizes $L_{A}=5$ and $L_{B}=5$, and (c) $N=2L=24$ in the even-parity sector with subsystem sizes $L_{A}=5$ and $L_{B}=6$. The horizontal black dot--dashed line corresponds to the Haar random values given by Eq.~\eqref{seq:von-analytical}. We consider $2^{20-L}$ Hamiltonian realizations. The inset shows the instantaneous slope $\Delta S_{A}^{\rm (vN)}(t)$, defined as $\Delta S_{A}^{\rm (vN)}((t_i-t_{i-1})/2)=(S_{A}^{\rm (vN)}(t_i)-S_{A}^{\rm (vN)}(t_{i-1}))/(t_i-t_{i-1})$, plotted as a function of $1/t$. Filled markers indicate the approximately linear regime used for fitting, while the dot-dashed lines show linear extrapolations based on ten selected points. The extrapolated intercept at $1/t\rightarrow 0$ provides the asymptotic entropy production rate.}  
    \label{sfig:usual_SYK_initial_state}
\end{figure}
For the usual SYK model, the Hamiltonian is given by \cite{Maldacena:2016hyu}
\begin{equation}
H_{\rm SYK}=\sqrt{\frac{6}{N^3}}\sum_{1\leq i<j<k<l\leq N}J_{ijkl}\psi_{i}\psi_{j}\psi_{k}\psi_{l}    
\end{equation}
where $\psi_i$ are the Fermionic operators which satisfy the Clifford algebra: $\{\psi_i,\psi_j\}=\delta_{ij}$ and $J_{ijkl}$ are standard Gaussian random variable with zero mean and unit variance. For brevity we use $H_{\text{SYK}}$ to denote the Hamiltonian of the usual SYK model with $q=4$.

Fig.~\ref{sfig:usual_SYK_initial_state} illustrates the time evolution of the average von-Neumann EE in the usual SYK model for several distinct initial states. The usual SYK exhibits the Gaussian orthogonal ensemble (GOE) for $N\equiv 0 \rm   \ {mod} \ 8 $, Gaussian unitary ensemble (GUE) for $N\equiv 2, \ 6 \rm   \ {mod} \ 8 $ and Gaussian symplectic ensemble (GSE) for $N\equiv 4 \rm   \ {mod} \ 8 $. Panels (a)-(c) correspond to system sizes $N=20$ (GSE), $N=22$ (GUE), and $N=24$ (GOE) respectively. For all system sizes considered here, th average von-Neumann EE increases rapidly from its initial value and gradually saturated at long times. A notable feature is the complete overlap of the curves associated with different initial states throughout the entire time evolution, indicating that the entanglement dynamics is essentially independent of the specific choice of the initial state. The distinction in the initial state dependence between the spin SYK-4 and usual SYK model can be attributed to the operator structure of the two models \cite{Pathak:2025udi}. In the usual SYK model, the Majorana fermion operators correspond to highly non-local strings of Pauli operators after the Jordan-Wigner transformation, leading to more efficient scrambling and state-independent growth dynamics. In contrast, the spin SYK-$q$ model is constructed directly from local spin operators, resulting in weaker scrambling and enhanced sensitivity to the choice of the initial state. We further perform the same refined analysis for the usual SYK model by examining the instantaneous entanglement-growth rate. We also observed the linear dependence of instantaneous slope on $1/t$ for all the system sizes considered. While the linear scaling is universal, the extrapolated values at the $t=\infty$ limit differ across system sizes, indicating that the entanglement growth rate depends on the system size. Therefore, while all system sizes exhibit the same qualitative behavior, the rate at which entanglement develops during the growth regime varies quantitatively with $N$.

\section{Spread of entanglement: Pure state}\label{ssec:eetgenstate}
In this section, we investigate pure-state entanglement measures, namely the von-Neumann entanglement entropy and the second-order Rényi entropy, for different values of $q$. We analyze their dynamics across various system sizes $L$ and for several distinct initial states.
\subsection{Random state}
\begin{figure}[htbp]
	\centering
	\includegraphics[width=16cm]{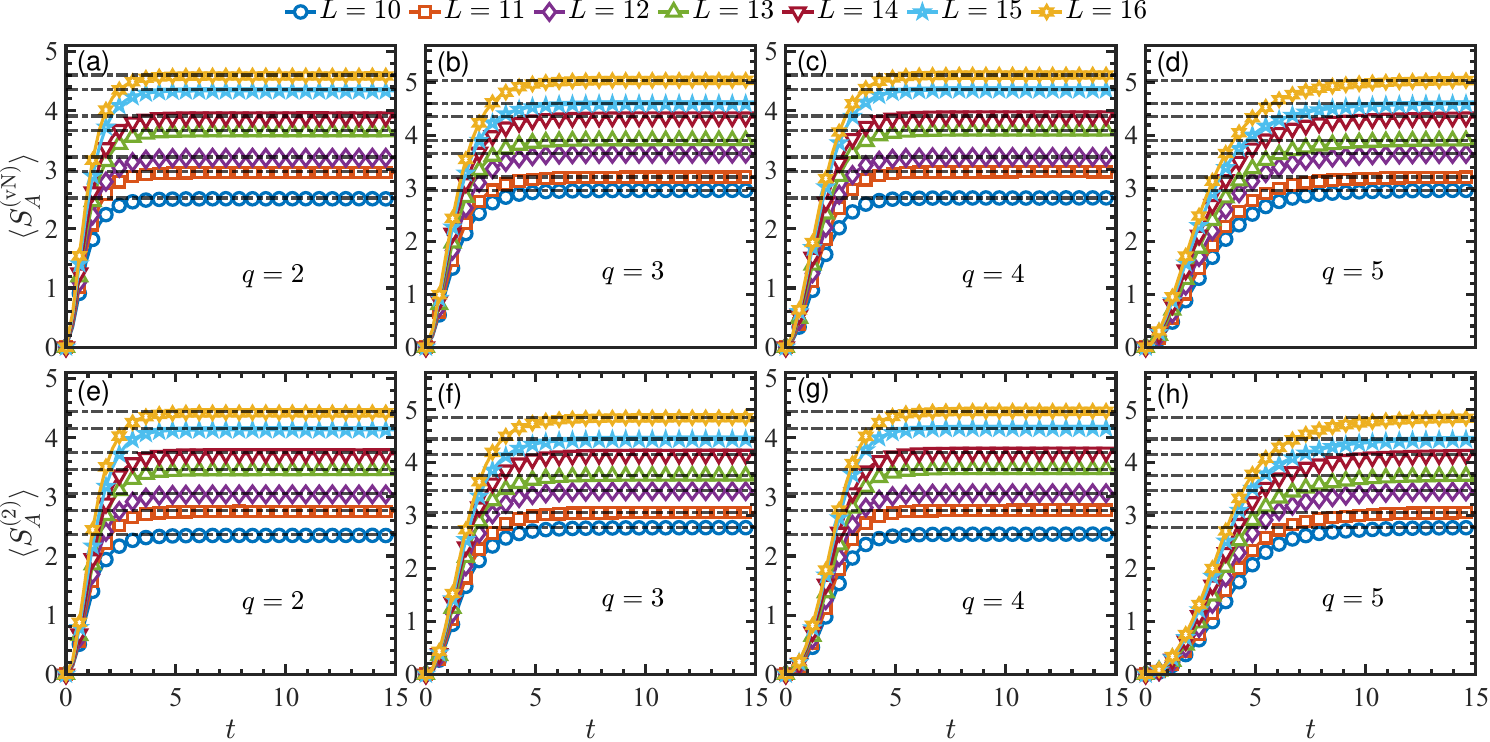}
	\caption{Dynamics of the average von-Neumann entanglement entropy $\langle S_{A}^{\rm (vN)}\rangle$ (upper panel) and second-order R\'enyi entropy $\langle S_{A}^{(2)}\rangle$ for spin SYK-$q$ model with different values of $q$ for different system sizes $L=10$--$16$. The initial state is given by Eq.~\eqref{seq:pstate}, where the values of $\theta_i$ and $\phi_i$ are chosen randomly. For even values of $q$, we simulate in the even-parity sector and for odd values of $q$, we consider the full Hilbert space. For even $q$, the subsystem sizes are chosen as $L_A=L_B=(L-1)/2$ for odd $L$, while for even $L$, we take $L_A=(L-2)/2$ and $L_B=L/2$. For odd $q$, the subsystem sizes are considered as $L_A=L_B=L/2$ for even $L$, while for odd $L$, we take $L_A=(L-1)/2$ and $L_B=(L+1)/2$. The horizontal black dot--dashed line corresponds to the Haar random values given by Eq.~\eqref{seq:von-analytical} (upper panel) and \eqref{seq:renyi-analytical} (lower panel). For each system size, we consider $2^{20-L}$ Hamiltonian realizations.}
    \label{sfig:ee_diff_q_random}
\end{figure}
Fig.~\ref{sfig:ee_diff_q_random} shows the average von-Neumann EE (upper panel) and 2-RE (lower panel) for the spin SYK-$q$ model initialized in the random product state defined in Eq.~\eqref{seq:pstate}. Results are presented for interaction orders $q=2,3,4,$ and $5$ and for several system sizes $L$. For all values of $q$, both entanglement measures exhibit an initial growth followed by saturation at long times. For a fixed interaction order $q$, the growth becomes progressively steeper as the system size increases, indicating a faster buildup of entanglement in larger systems. The same trend is observed across all values of $q$ considered here. At late times the numerical results remain close to the Haar random values, for different system sizes $L$ shown by the horizontal dashed lines. While the overall qualitative behavior is similar for all interaction orders, the rate of entanglement growth depends on $q$ for a fixed system size, with larger values of $q$ generally leading to a slower approach toward the saturation regime. A similar trend is observed for the second-order R\'enyi entropy.

\pagebreak
\subsection{Superposition state}
\begin{figure}[htbp]
	\centering
	\includegraphics[width=16cm]{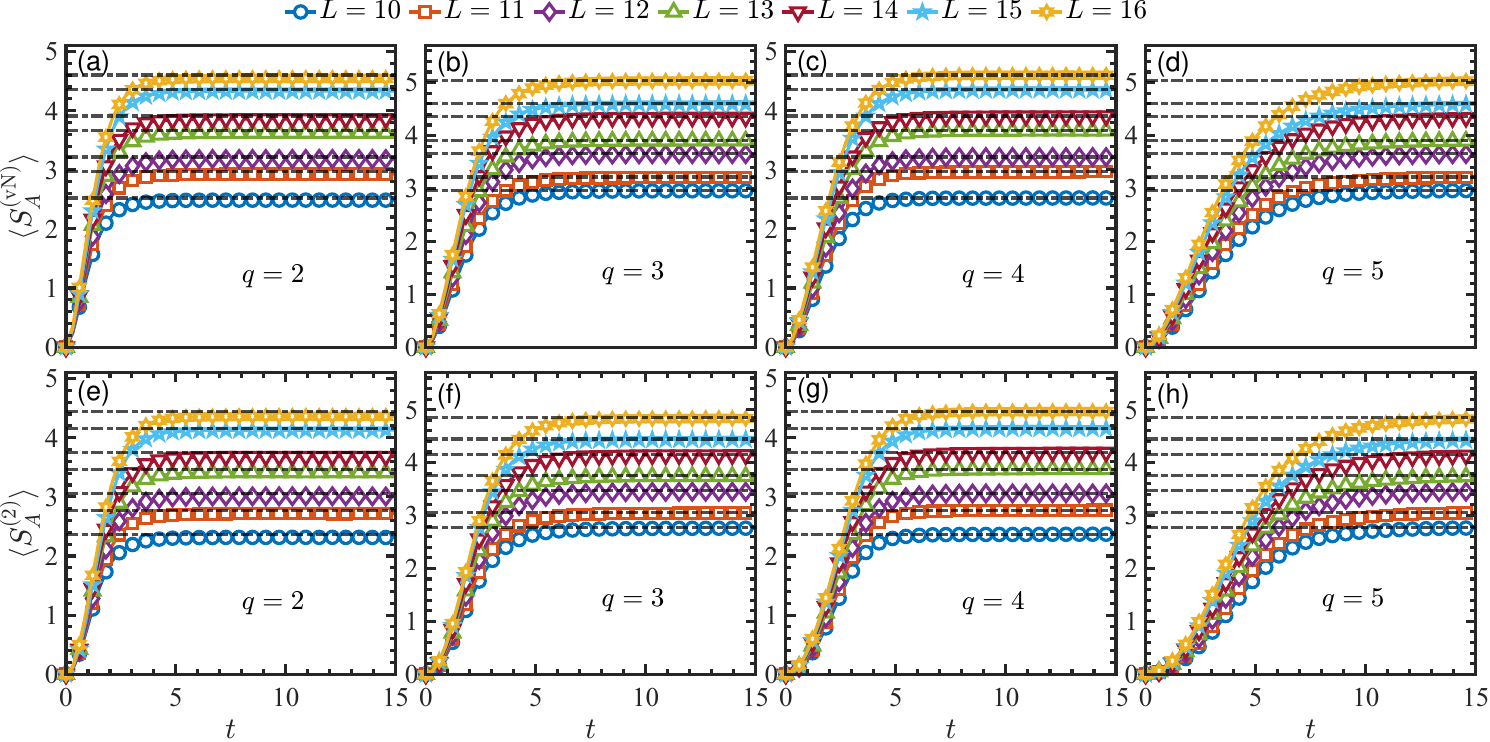}
	\caption{Dynamics of the average von-Neumann entanglement entropy $\langle S_{A}^{\rm (vN)}\rangle$ (upper panel) and second-order R\'enyi entropy $\langle S_{A}^{(2)}\rangle$ for spin SYK-$q$ model with different values of $q$ for different system sizes $L=10$--$16$. The initial state is chosen as the symmetric superposition state corresponding to $\theta_i=\pi/2$ and $\phi_i=0$ for all sites in Eq.~\eqref{seq:pstate}. For even values of $q$, we simulate in the even-parity sector and for odd values of $q$, we consider the full Hilbert space. For even $q$, the subsystem sizes are chosen as $L_A=L_B=(L-1)/2$ for odd $L$, while for even $L$, we take $L_A=(L-2)/2$ and $L_B=L/2$. For odd $q$, the subsystem sizes are considered as $L_A=L_B=L/2$ for even $L$, while for odd $L$, we take $L_A=(L-1)/2$ and $L_B=(L+1)/2$. The horizontal black dot--dashed line corresponds to the Haar random values given by Eq.~\eqref{seq:von-analytical} (upper panel) and \eqref{seq:renyi-analytical} (lower panel). For each system size, we consider $2^{20-L}$ Hamiltonian realizations.}
    \label{sfig:ee_diff_q_superposition}
\end{figure}

Fig.~\ref{sfig:ee_diff_q_superposition} shows the averaged von-Neumann EE (upper panel) and 2-RE (lower panel) for the spin SYK-$q$ model initialized in the symmetric superposition state. The symmetric superposition state represents a homogeneous and highly coherent initial configuration. Since it is initially unentangled, it provides a useful benchmark for examining how local quantum coherence \cite{Baumgratz:2014} is converted into many-body entanglement during the subsequent dynamics. Results are presented for interaction order $q=2,3,4$ and $5$ and for several system sizes $L$. The entanglement dynamics for the symmetric superposition state exhibit trends similar to those observed for the random product state. For all values of $q$, both the von-Neumann and second-order R\'enyi entropies show an initial growth followed by saturation at long times. For a fixed $q$, the growth becomes steeper with increasing system size, whereas increasing $q$ leads to a slower buildup of entanglement. The numerical results remain close to the Haar random values at late times.
\clearpage
\subsection{Other states of experimental interest}
To examine the robustness of the scrambling and entanglement dynamics discussed in the main text, we consider several experimentally relevant initial product states. These states possess distinct spin-ordering patterns and provide a useful platform for investigating how the initial configuration influences the generation and propagation of entanglement. Specifically, we analyze the dynamics starting from the N\'eel state, $|\psi_1\rangle = |\uparrow \downarrow \uparrow \downarrow \cdots \rangle$, the double N\'eel state $|\psi_2\rangle = |\uparrow \uparrow \downarrow \downarrow \uparrow \uparrow\cdots \rangle$, the single excitation state $|\psi_3\rangle = |\uparrow \downarrow \downarrow \downarrow \downarrow \cdots \rangle$, and the domain-wall state $|\psi_4\rangle = |\uparrow \uparrow \uparrow \cdots \downarrow \downarrow  \downarrow \rangle$. For each initial state, we compute the pure state entanglement measures such as von-Neumann entanglement entropy and second-order R\'enyi entropy and compare their time evolution across different system size $L$ and different interaction order $q$. 

\subsubsection{N\'eel state}
\begin{figure}[htbp]
	\centering
	\includegraphics[width=16cm]{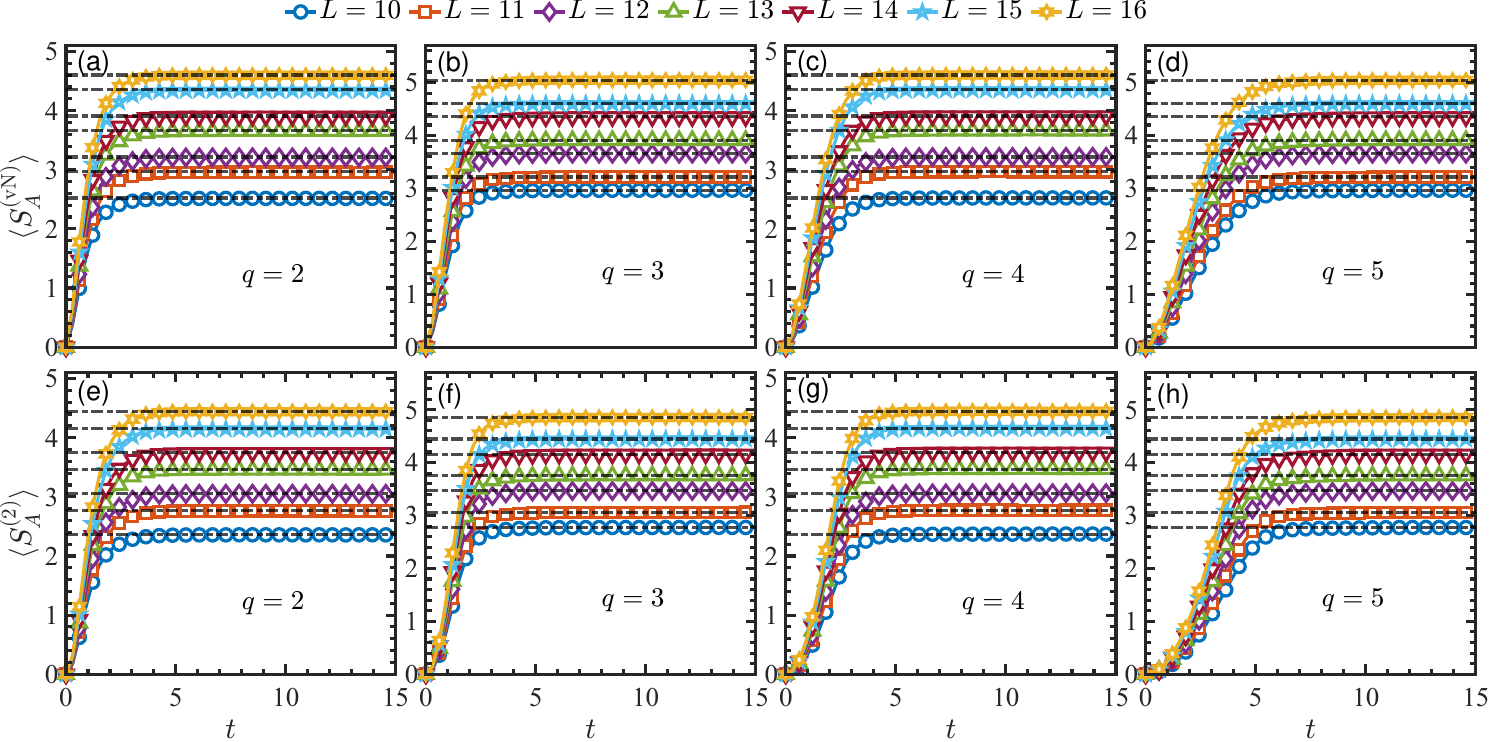}
	\caption{Dynamics of the average von-Neumann entanglement entropy $\langle S_{A}^{\rm (vN)}\rangle$ (upper panel) and second-order R\'enyi entropy $\langle S_{A}^{(2)}\rangle$ for spin SYK-$q$ model with different values of $q$ for different system sizes $L=10$--$16$. The initial state is chosen as the N\'eel state. For even values of $q$, we consider only the even-parity sector and for odd values of $q$, we consider the full Hilbert space. For even $q$, the subsystem sizes are chosen as $L_A=L_B=(L-1)/2$ for odd $L$, while for even $L$, we take $L_A=(L-2)/2$ and $L_B=L/2$. For odd $q$, the subsystem sizes are considered as $L_A=L_B=L/2$ for even $L$, while for odd $L$, we take $L_A=(L-1)/2$ and $L_B=(L+1)/2$. The horizontal black dot--dashed line corresponds to the Haar random values given by Eq.~\eqref{seq:von-analytical} (upper panel) and \eqref{seq:renyi-analytical} (lower panel). For each system size, we consider $2^{20-L}$ Hamiltonian realizations.}
    \label{sfig:ee_diff_q_neel}
\end{figure}

The N\'eel state $|\psi_1\rangle = |\uparrow \downarrow \uparrow \downarrow \cdots \rangle$ is one of the most widely studied initial states in quantum simulators and many-body spin systems \cite{Bloch:2015,Abanin:2019,Monroe:2021,Bernien:2017,Smith:2016}. This state is experimentally relevant owing to its simple staggered magnetic order and has been extensively studied as an initial condition in studies of non-equilibrium spin dynamics. Fig.~\ref{sfig:ee_diff_q_neel} shows the evoluation of the average von-Neumann EE and 2-RE for different interaction order $q$ and several system sizes $L$. The entanglement dynamics exhibits trends similar to those observed for the random and symmetric superposition states. For all values of $q$, both entropies display an initial growth followed by saturation at long times. Furthermore, increasing $q$ results in a slower buildup of entanglement, while the saturation values remain close to the corresponding Haar-random predictions.  

\clearpage

\subsubsection{Double N\'eel state}
\begin{figure}[htbp]
	\centering
	\includegraphics[width=16cm]{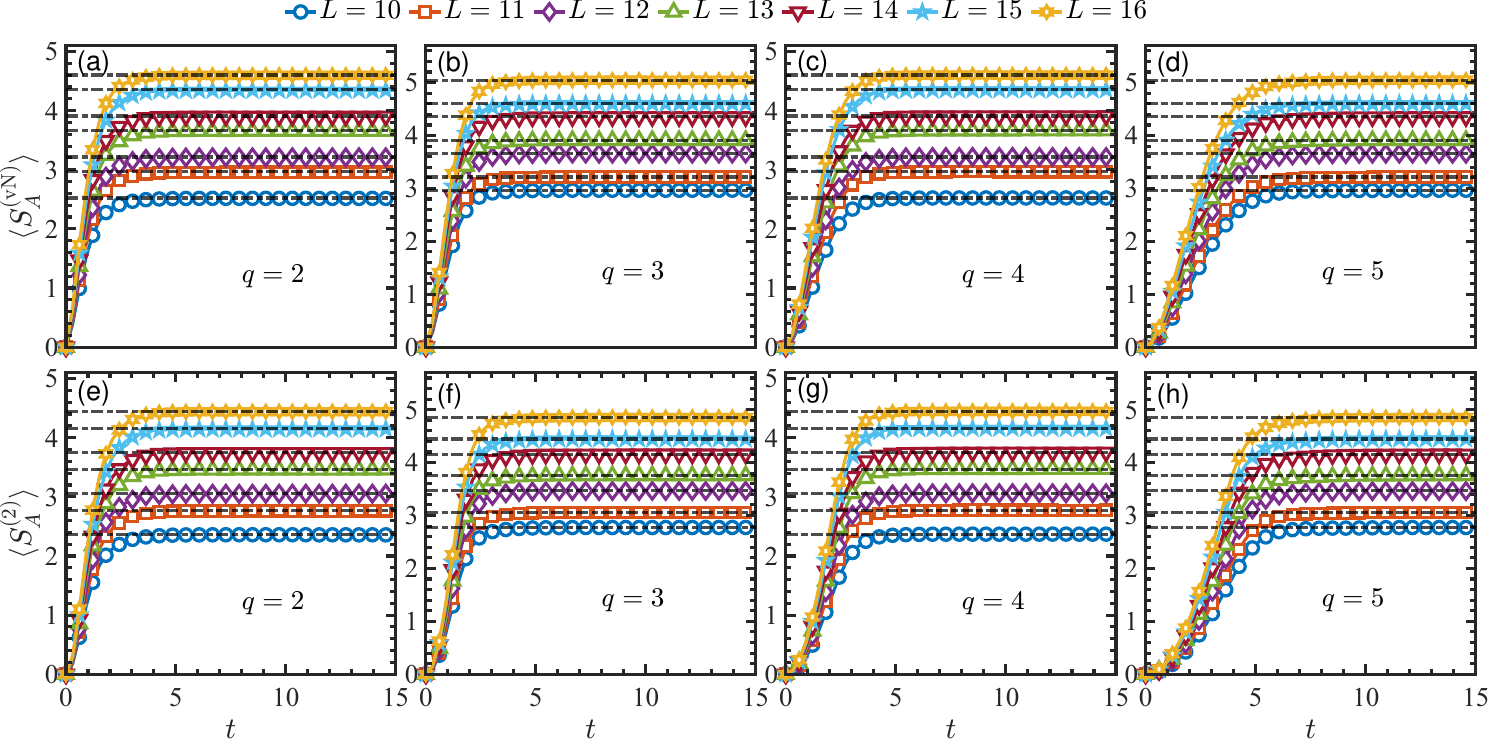}
	\caption{Dynamics of the average von-Neumann EE (upper panel) and 2-RE for spin SYK-$q$ model with different values of $q$ for different system sizes $L=10$--$16$. The initial state is chosen as the double N\'eel state. For even values of $q$, we consider only the even-parity sector and for odd values of $q$, we consider the full Hilbert space. For even $q$, the subsystem sizes are chosen as $L_A=L_B=(L-1)/2$ for odd $L$, while for even $L$, we take $L_A=(L-2)/2$ and $L_B=L/2$. For odd $q$, the subsystem sizes are considered as $L_A=L_B=L/2$ for even $L$, while for odd $L$, we take $L_A=(L-1)/2$ and $L_B=(L+1)/2$. The horizontal black dot--dashed line corresponds to the Haar random values given by Eq.~\eqref{seq:von-analytical} (upper panel) and \eqref{seq:renyi-analytical} (lower panel). For each system size, we consider $2^{20-L}$ Hamiltonian realizations.}
    \label{sfig:ee_diff_q_doubleneel}
\end{figure}

We now consider the double-N\'eel state. Compared with the N\'eel state, this state is another experimentally relevant product state characterized by a period-four spin pattern. Fig.~\ref{sfig:ee_diff_q_doubleneel} shows the von-Neumann and second-order R\'enyi ntanglement entropies for interaction orders $q=2,3,4,$ and $5$ with several system sizes $L$. The entanglement dynamics exhibit trends similar to those observed for the random, symmetric superposition, and N\'eel states. For all values of $q$, both entropies display an initial growth followed by saturation at long times. For a fixed interaction order $q$, the growth becomes progressively steeper with increasing system size, indicating faster entanglement generation in large system sizes. On the other hand, increasing $q$ leads to a slower buildup of entanglement. At late times, the numerical results remain close to the corresponding Haar predictions.

\clearpage
\subsubsection{Single excitation state}
\begin{figure}[htbp]
	\centering
	\includegraphics[width=16cm]{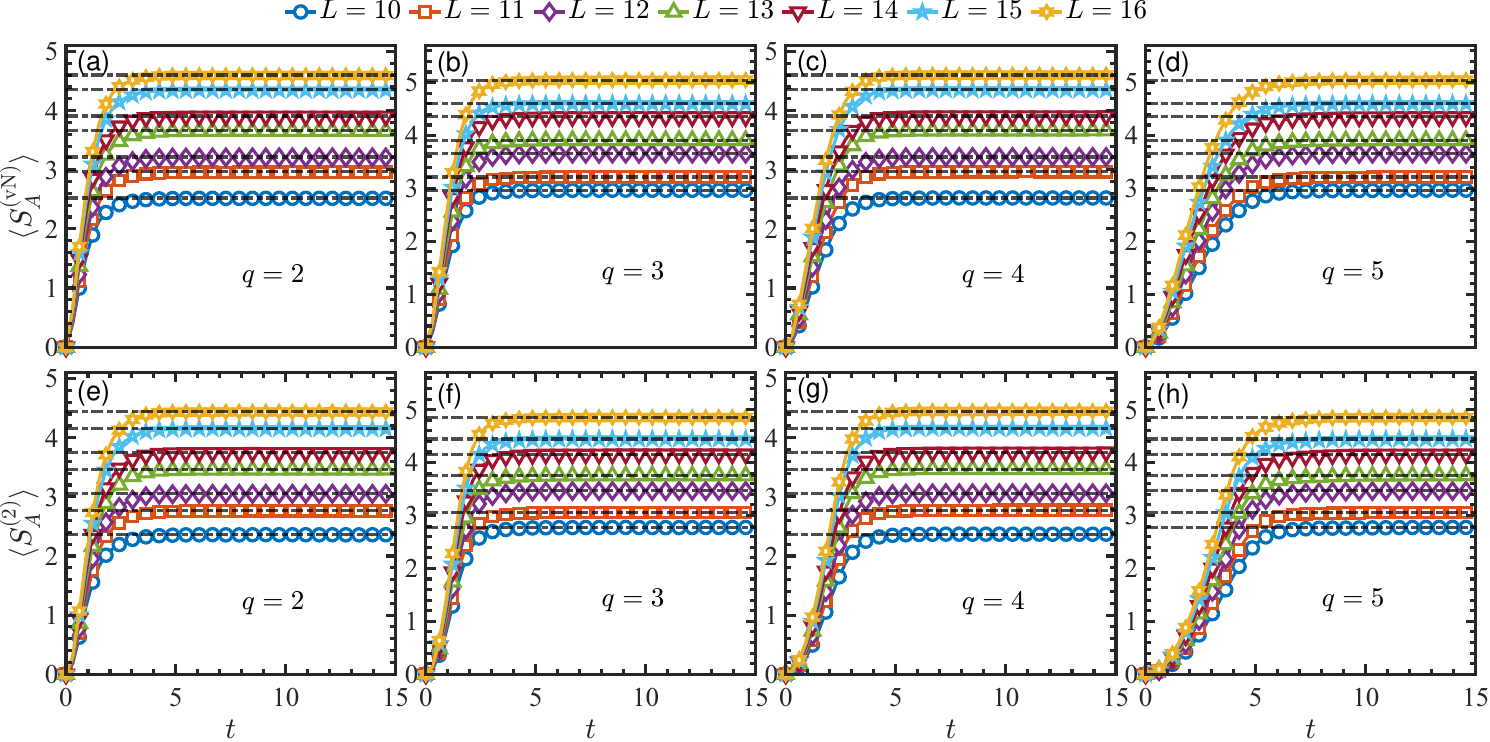}
	\caption{Dynamics of the average von-Neumann EE (upper panel) and 2-RE for spin SYK-$q$ model with different values of $q$ for different system sizes $L=10$--$16$. The initial state is chosen as the single excitation state $|\uparrow \downarrow \downarrow \downarrow \cdots \rangle$. For even values of $q$, we consider the even-parity sector and for odd values of $q$, we consider the full Hilbert space. For even $q$, the subsystem sizes are chosen as $L_A=L_B=(L-1)/2$ for odd $L$, while for even $L$, we take $L_A=(L-2)/2$ and $L_B=L/2$. For odd $q$, the subsystem sizes are considered as $L_A=L_B=L/2$ for even $L$, while for odd $L$, we take $L_A=(L-1)/2$ and $L_B=(L+1)/2$. The horizontal black dot--dashed line corresponds to the Haar random values given by Eq.~\eqref{seq:von-analytical} (upper panel) and \eqref{seq:renyi-analytical} (lower panel). For each system size, we consider $2^{20-L}$ Hamiltonian realizations.}
    \label{sfig:ee_diff_q_singleexc}
\end{figure}

We next consider the single-excitation state $|\psi_3\rangle=|\uparrow \downarrow \downarrow \downarrow \downarrow \cdots \rangle$, corresponding to $\theta_1=0$ and $\theta_i=\pi$  ea $i=2,3,\ldots,L$, with $\phi_i=0$ for all sites in Eq.~\eqref{seq:pstate}. This state contains a single spin excitation embedded in an otherwise polarized background and is of interest for studying the dynamics generated by a localized spin perturbation \cite{Jurcevic:2014,Monroe:2017,Monroe:2021rmp}. This state is also serves as another experimentally relevant product state. Fig.~\ref{sfig:ee_diff_q_singleexc} shows the average von-Neumann EE and 2-RE for different interaction order $q$ and different system sizes $L$. The entanglement dynamics exhibits trends similar to those previously observed initial states. For all values of $q$, both entropies display an initial growth followed by saturation at long times. For a fixed interaction order $q$, the growth becomes progressively steeper with increasing system size, indicating faster entanglement generation. On the other hand, increasing $q$ leads to a slower buildup of entanglement. At late times, the numerical results remain close to the corresponding Haar predictions. 

\clearpage

\subsubsection{Domain-wall state}
\begin{figure}[htbp]
	\centering
	\includegraphics[width=16cm]{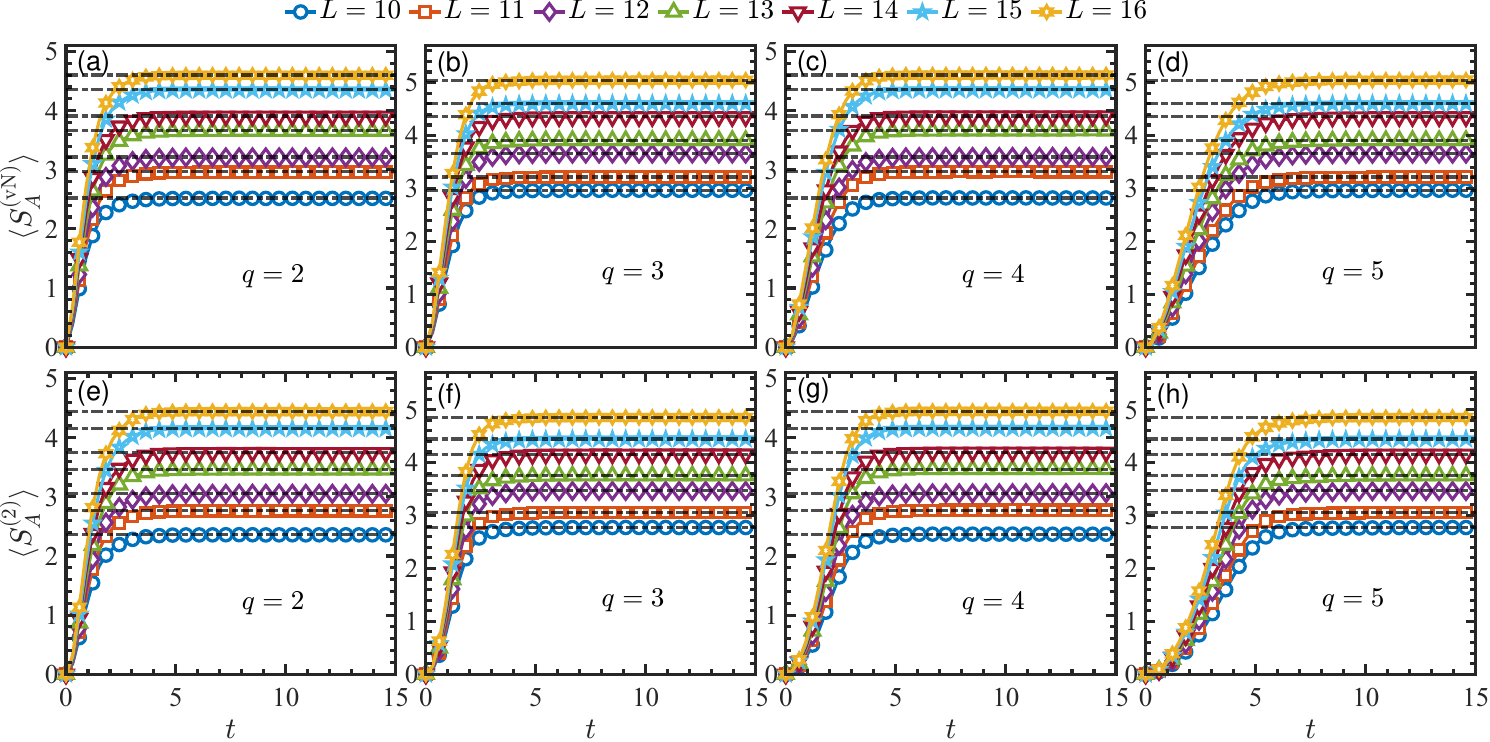}
	\caption{Dynamics of the average von-Neumann EE (upper panel) and 2-RE for spin SYK-$q$ model with different values of $q$ for different system sizes $L=10$--$16$. The initial state is chosen as the domain-wall product state $|\uparrow \uparrow\uparrow \cdots \downarrow \downarrow  \downarrow \rangle$. For even values of $q$, we simulate in the even-parity sector and for odd values of $q$, we consider the full Hilbert space. For even $q$, the subsystem sizes are chosen as $L_A=L_B=(L-1)/2$ for odd $L$, while for even $L$, we take $L_A=(L-2)/2$ and $L_B=L/2$. For odd $q$, the subsystem sizes are considered as $L_A=L_B=L/2$ for even $L$, while for odd $L$, we take $L_A=(L-1)/2$ and $L_B=(L+1)/2$. The horizontal black dot--dashed line corresponds to the Haar random values given by Eq.~\eqref{seq:von-analytical} (upper panel) and \eqref{seq:renyi-analytical} (lower panel). For each system size, we consider $2^{20-L}$ Hamiltonian realizations.}
    \label{sfig:ee_diff_q_domainwall}
\end{figure}

Finally, we consider the domain-wall state $|\psi_4\rangle = |\uparrow \uparrow \uparrow \cdots \downarrow \downarrow  \downarrow \rangle$. This state is characterized by a sharp interface separating two oppositely polarized domains and represents a strongly inhomogeneous initial configuration. Domain-wall states are widely employed in studies of non-equilibrium quantum dynamics, transport and relaxation processes \cite{Ljubotina:2017,Misguich:2017}. Fig.~\ref{sfig:ee_diff_q_domainwall} shows the dynamics of the average von-Neumann EE and 2-RE for interaction orders $q=2,3,4,$ and $5$ and different system sizes $L$. The entanglement dynamics exhibits trends similar to those previously observed initial states. For all values of $q$, both entropies display an initial growth followed by saturation at long times. For a fixed interaction order $q$, the growth becomes progressively steeper with increasing system size, indicating faster entanglement generation. On the other hand, increasing $q$ leads to a slower buildup of entanglement. At late times, the numerical results remain close to the corresponding Haar predictions. 

Remarkably, for fixed $q$ and $L$, the entanglement dynamics of the N\'eel, double-N\'eel, single excitation, and domain-wall states are nearly indistinguishable despite their markedly different spin configurations. This observation points to the robustness of the scrambling dynamics in the spin SYK-$q$ model. Owing to the all-to-all nature of the interactions, local spatial structures present in the initial state are rapidly mixed, leading to a universal entanglement growth and eventual saturation toward Haar-random values.  

\clearpage
\section{Spread of entanglement: Mixed state}\label{ssec:mixedstate}
In this section, we investigate mixed-state entanglement measures, namely the R\'enyi-1/2 mutual information $(I_{A:B}^{(1/2)}(t))$, entanglement negativity $(2\mathcal{E}(t))$, and odd entropy $(\frac{2}{3}\mathcal{E}^{(o)}(t))$, for spin SYK-$q$ model with different values of $q$. We analyze their dynamics across various small system sizes $L$. For the mixed state entanglement measure, we choose the initial state corresponding to Eq.~\eqref{seq:pstate}, where the values of $\theta_{i}$ and $\phi_{i}$ are chosen randomly.

\begin{figure}[bp]
	\centering
	\includegraphics[width=9cm]{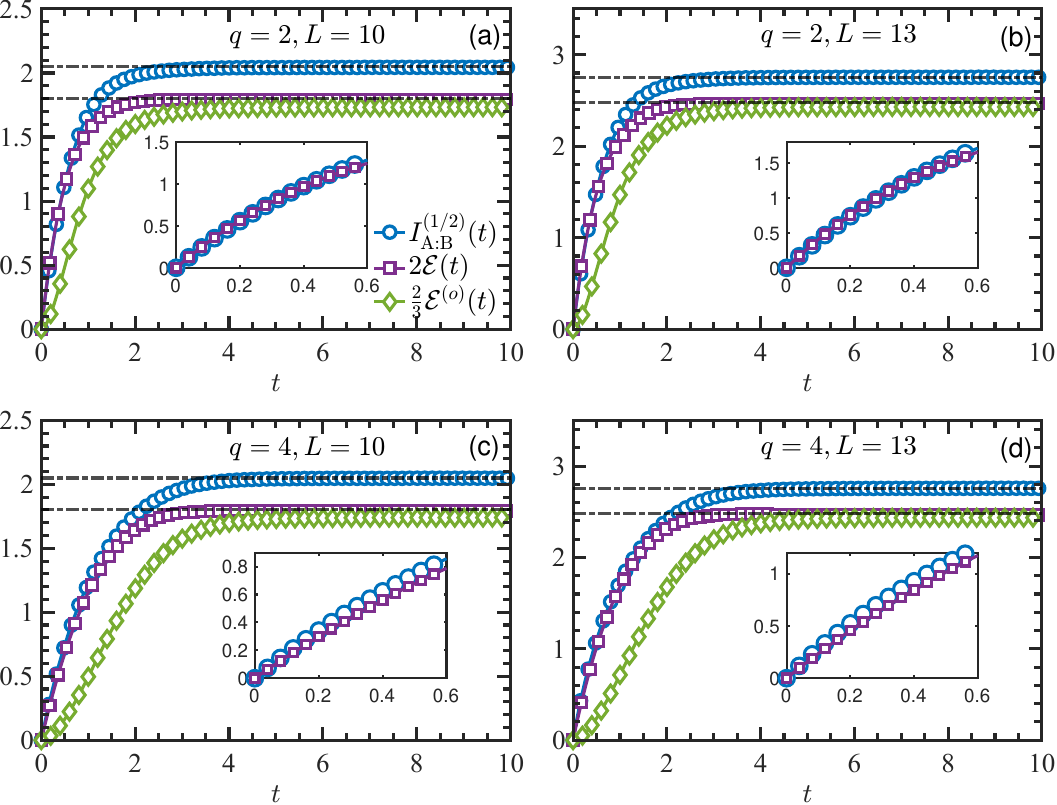}
	\caption{The dynamics of R\'enyi-1/2 mutual information $(I_{A:B}^{(1/2)}(t))$, entanglement negativity $(2\mathcal{E}(t))$ and odd entropy $(\frac{2}{3}\mathcal{E}^{(o)}(t))$ for spin SYK-$q$ model with (a) $q=2,L=10$, (b) $q=2,L=13$, (c) $q=4,L=10$, (d) $q=4,L=13$ and initial state is given by Eq.~\eqref{seq:pstate}. The values of $\theta_{i}$ and $\phi_{i}$ are chosen randomly. The horizontal black dot-dashed lines indicate to the Haar random values. The inset shows the zoomed view of $(I_{A:B}^{(1/2)}(t))$ and $(2\mathcal{E}(t))$ in the early time regime for $q=2$ and $q=4$ with system sizes $L=10$ and $L=13$. We consider $2^{20-L}$ Hamiltonian realizations. For $q=2$ and $q=4$, we consider even parity sector of system size $L=10$ with equal subsystem sizes $L_A=L_B=L_C=3$, while for system size $L=13$, we consider subsystem sizes $L_A=L_B=L_C=4$.}
    \label{sfig:entropy_even_q_mixed_state}
\end{figure}

\begin{figure}[htbp]
	\centering
	\includegraphics[width=9cm]{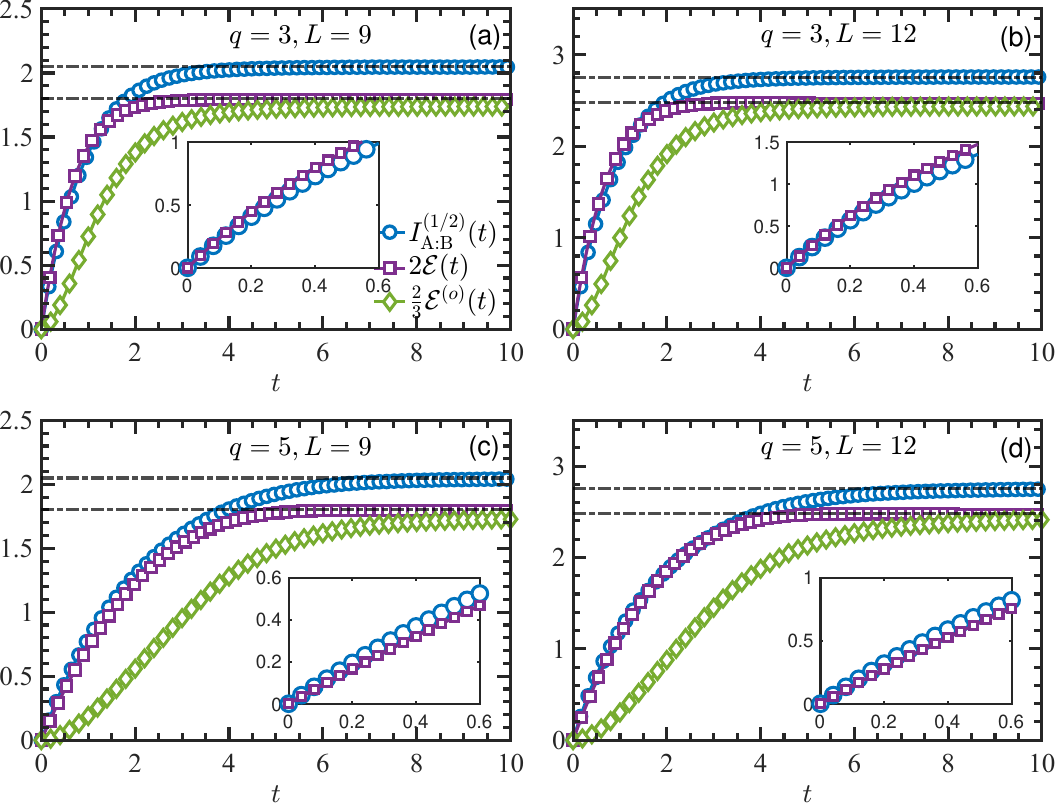}
	\caption{The dynamics of R\'enyi-1/2 mutual information $(I_{A:B}^{(1/2)}(t))$, entanglement negativity $(2\mathcal{E}(t))$ and odd entropy $(\frac{2}{3}\mathcal{E}^{(o)}(t))$ for spin SYK-$q$ model with (a) $q=3,L=9$, (b) $q=3,L=12$, (c) $q=5,L=9$, (d) $q=5,L=12$ and initial state is given by Eq.~\eqref{seq:pstate}. The value of $\theta_{i}$ and $\phi_{i}$ are chosen randomly. The horizontal black dot-dashed lines indicate to the Haar random values. The inset shows the zoomed view of $(I_{A:B}^{(1/2)}(t))$ and $(2\mathcal{E}(t))$ in the early time regime for $q=3$ and $q=5$ with system sizes $L=9$ and $L=12$. We consider $2^{20-L}$ Hamiltonian realizations. For $q=3$ and $q=5$, we consider full-Hilbert space of system size $L=10$ with equal subsystem sizes $L_A=L_B=L_C=3$, while for system size $L=13$, we consider subsystem sizes $L_A=L_B=L_C=4$.}
    \label{sfig:entropy_odd_q_mixed_state}
\end{figure}

Figs.~\ref{sfig:entropy_even_q_mixed_state} and \ref{sfig:entropy_odd_q_mixed_state} show the dynamics of the R\'enyi-1/2 mutual information $(I_{A:B}^{(1/2)}(t))$, entanglement negativity $(2\mathcal{E}(t))$, and odd entropy $(\frac{2}{3}\mathcal{E}^{(o)}(t))$ for random product initial states with small system sizes. Fig.~\ref{sfig:entropy_even_q_mixed_state} shows the results for the even-$q$ values ($q=2,4$), while
Fig.~\ref{sfig:entropy_odd_q_mixed_state} displays the corresponding results for the odd-$q$ values ($q=3,5$). The qualitative behavior closely follows that observed in the main text. In all cases, the three mixed state entanglement measures exhibit a rapid initial growth associated with the buildup of quantum correlations, followed by saturation to values consistent with Haar prediction. For $q=2$ case, the R\'enyi-1/2 mutual information and the entanglement negativity are nearly indistinguishable throughout the entire growth regime, as highlighted in the insets even for small system sizes. This provides further numerical evidence for the relation $$I_{A:B}^{(1/2)}(t)=2\mathcal{E}(t)$$ The qualitative agreement with the larger-system results presented in the main text indicates that the observed relation is robust against finite-size effects. The agreement remains partially valid for $q=3$, although small deviations become visible as the correlations grow. For higher interaction orders, namely $q=4$ and $q=5$, the two quantities clearly show systematic deviation in the early growth regime, indicating a progressive breakdown of this correspondence as the dynamics become increasingly dominated by higher-body interactions. Nevertheless, both quantities eventually saturate close to their respective Haar-random values.  

A notable feature common to both the even and odd values of $q$ is the behavior of the odd entropy. While $\frac{2}{3}\mathcal{E}^{(o)}(t)$ grows more slowly than $I_{A:B}^{(1/2)}(t)$ and $2\mathcal{E}(t)$ during the transient regime, it approaches the entanglement negativity at late times. The convergence of these two mixed-state entanglement measures becomes increasingly pronounced near saturation, where both attain values compatible with the Haar prediction. Comparing different interaction orders further reveals that increasing $q$ systematically slows down the buildup of quantum correlations and consistent with the ordering $v_2>v_3\gtrsim v_4>v_5$ in natural time scale and $\tilde v_2< \tilde v_3<\tilde v_4\lesssim \tilde v_5$ in energy fixed time scale. The persistence of this ordering for smaller system sizes demonstrates that the dependence of scrambling dynamics on the interaction order is robust and not an artifact of finite-size effects. 

\section{$p$-spin model}\label{ssec:pspin}
To further examine the role of multi-body interactions in entanglement growth, we consider a generic all-to-all interacting $p$-spin model, 
\begin{equation}
H_{p}
=\sum_{i_1<\cdots<i_p}^{L}J_{i_1 \cdots i_p}\prod_{k=1}^{p}\sigma_{i_k}^{\alpha_{i_k}},\quad
\alpha_{i_k} \in \{x,y,z\}.
\label{seq:pbody}
\end{equation}
where each spin component $\alpha_{i_k}$ is independently chosen from $\{x, y,z\}$ and the couplings $J_{i_1 \cdots i_p}$ standard Gaussian random variable with zero mean and standard deviation to be unity. $\sigma^{x,y,z}$ denotes the standard Pauli matrices. This model serves as a minimal and generic realization of all-to-all $p$-body local interactions. It is important to emphasize that its operator structure differs substantially from that of the spin SYK-$q$ model. In spin SYK-$q$, a given lattice site can appear through different spin components (e.g., both $\sigma^{x}$ and $\sigma^{y}$) within the same interaction term, generating a highly structured operator ensemble. In contrast, the present model associates a single spin operator with each participating site, resulting in a simpler and more generic interaction pattern. Thus, the agreement between the two models is nontrivial and indicates that the observed behavior is not tied to the specific operator structure of the spin SYK-$q$ model. 
\begin{figure}[htbp]
	\centering
	\includegraphics[width=12cm]{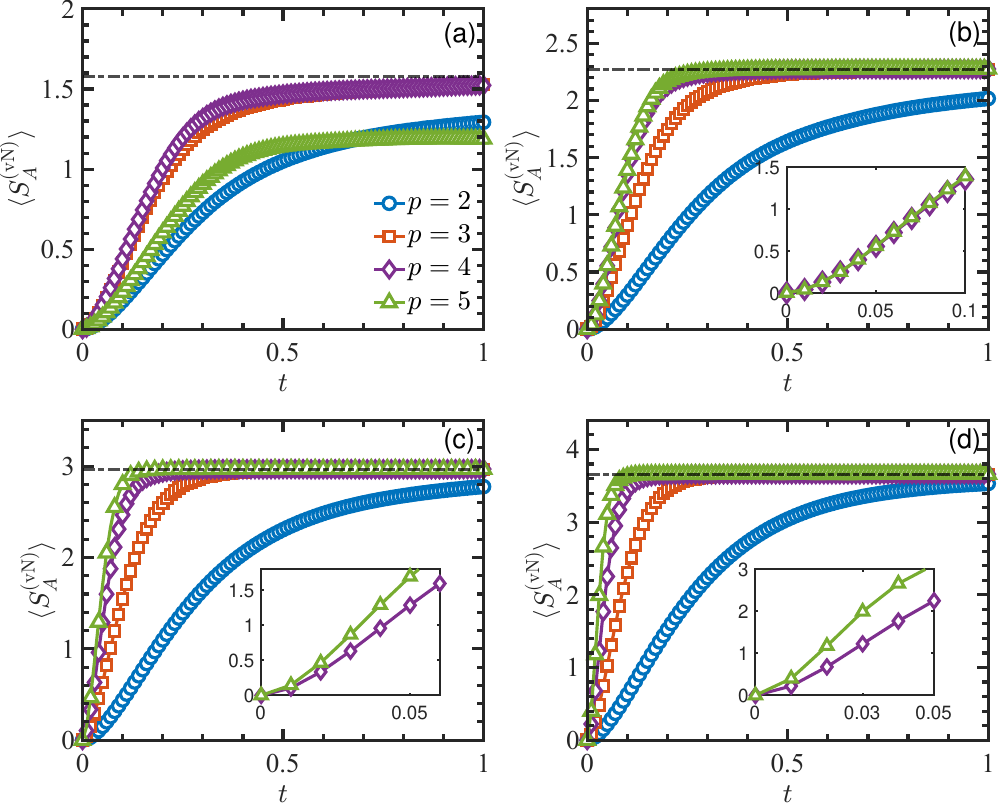}
	\caption{The dynamics of the average von-Neumann EE for different values of $p$ with small system sizes $L$: (a) $L=6$, (b) $L=8$, (c) $L=10$, and (d) $L=12$, for the $p$-spin interaction model described by Eq.~\eqref{seq:pbody}. We consider $2^{20-L}$ Hamiltonian realizations. The initial state is given by Eq.~\eqref{seq:pstate}, where the values of $\theta_i$ and $\phi_i$ are chosen randomly. The horizontal black dot--dashed line corresponds to the Haar random values given by Eq.~\eqref{seq:von-analytical}. }
    \label{sfig:ee_p_spin}    
\end{figure}

Fig.~\ref{sfig:ee_p_spin} shows the averaged von-Neumann EE dynamics for $p=2,3,4$ and $5$ with different system sizes. A common feature across all interaction orders is the rapid growth of entanglement from the initial state corresponding to Eq.~\eqref{seq:pstate}, where $\theta_{i}$ and $\phi_{i}$ are chosen randomly, followed by saturation at values close to the Haar prediction given by Eq.~\eqref{seq:von-analytical}. For smaller system sizes, the ordering of entanglement growth rates is not strictly monotonic in $p$. In particular, higher-body interactions do not always lead to faster entanglement production, reflecting the significant influence of finite-size effects and the competition between different all-to-all interaction orders. However, as the system size increases, the growth curves progressively reorganize and eventually approach the ordering $v_2<v_3<v_4<v_5$, where $v_p$ denotes the linear growth rate of the entanglement entropy. This trend is clearly visible in the larger system sizes ($L=10, 12$) as shown in Fig.~\ref{sfig:ee_p_spin}(c) and (d), where higher-body interactions consistently generate faster scrambling and earlier saturation to the Haar random value. The emergence of the same ordering observed in the spin SYK-$q$ model demonstrates that the phenomenon is remarkably robust and does not depend on the detailed operator structure of the Hamiltonian. Instead, it appears to be a generic consequence of increasing interaction order in sufficiently large all-to-all interacting spin systems.

\section{Genuine spin SYK-$q$ model}\label{ssec:gsyk}
The spin SYK-$q$ model contains interaction terms in which different spin components associated with the same lattice site can appear simultaneously within a single operator string. Such terms generate effective self-site contributions and lead to a highly structured operator ensemble. A genuine version of the model can be obtained by excluding all operator combinations involving multiple spin components acting on the same site. This corresponds to retaining only terms with $\eta_{i_1\cdots i_q}=0$, thereby eliminating all self-site contributions and ensuring that every interaction term acts on $q$ distinct lattice sites.

This construction is closely related to the generic all-to-all interacting $p$-spin model introduced in Eq.~\eqref{seq:pbody}. In both cases, each participating site appears only once within a given interaction term and is represented by a single local spin operator. Consequently, the genuine spin SYK-$q$ model can be viewed as a structured realization of genuine many-body interactions, whereas the $p$-spin model provides a more generic random realization of the same principle. Comparing the two different version of the spin SYK-$q$ model, therefore allows us to determine whether the entanglement dynamics observed in the main text originate from the special operator structure of spin SYK-$q$ or reflect a more universal feature of all-to-all interacting systems.

\begin{figure}[htbp]
	\centering
	\includegraphics[width=0.6\textwidth]{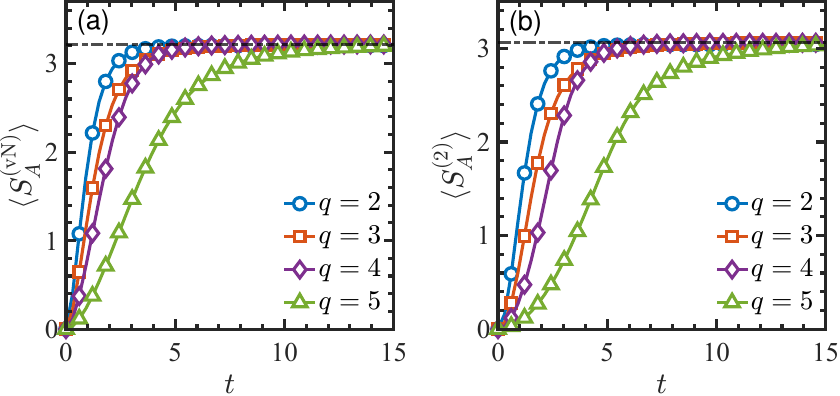}
	\caption{Evolution of (a) average von-Neumann EE and (b) 2-RE with time for genuine spin SYK-$q$ model with $q=2,3,4,5$ by setting $\eta_{i_1\cdots i_q}=0$. The initial state given by Eq. \eqref{seq:pstate}. The value of $\theta_{i}$ and $\phi_{i}$ are chosen randomly. We consider $L=12$, even parity block for even $q$ and $L=11$ with full Hilbert for odd $q$. The sub-systems sizes are $L_{A}=5, L_{B}=6$ for both the cases. The horizontal black dot--dashed line corresponds to the Haar random values given by Eq.~\eqref{seq:von-analytical} and \eqref{seq:renyi-analytical} respectively. We consider $2^{20-L}$ Hamiltonian realizations and initial state is changed over each realization.}
    \label{sfig:ee_genuine_SYK}
\end{figure}

Fig.~\ref{sfig:ee_genuine_SYK} shows the averaged von-Neumann entanglement entropy $\langle S_A^{(\mathrm{vN})}\rangle$ and second-order R\'enyi entropy $\langle S_A^{(2)}\rangle$ for the genuine spin SYK-$q$ model with $q=2,3,4$ and $5$. A common feature across all interaction orders is the rapid generation of entanglement from the initial random product state, followed by saturation at values close to those expected for Haar-random states. Both entanglement measures exhibit nearly identical qualitative behavior throughout the evolution, indicating that the scrambling dynamics are largely independent of the specific choice of entanglement measure.

Although the qualitative behavior is universal, the interaction order significantly affects the scrambling timescale. The $q=2$ case exhibits the fastest entanglement growth and reaches its saturation value at the earliest time, while increasing $q$ progressively slows the dynamics. As a result, the entanglement growth follows the ordering $v_2>v_3>v_4>v_5$ in natural time scale, where $v_{q}$ denotes the growth rate of the entanglement entropy. The similar ordering is also observed in the second-order R\'enyi entropy case. Most importantly, these results closely mirror those obtained for the spin SYK-$q$ model. Despite eliminating all operator strings containing multiple spin components acting on the same site through the constraint $\eta_{i_1\cdots i_q}=0$, the qualitative features of the dynamics remain unchanged: the entropies exhibit rapid growth, saturate near the Haar-random value, and display a systematic dependence of the scrambling timescale on the interaction order. The persistence of these features demonstrates that they do not originate from the specific operator structure of the spin SYK-$q$ model. Rather, they are robust consequences of all-to-all interacting many-body dynamics and remain intact even when only genuine $q$-body interactions are retained. 

\clearpage
\section{Density of states}\label{ssec:dos}
\begin{figure}[htbp]
	\centering
	\includegraphics[width=0.6\textwidth]{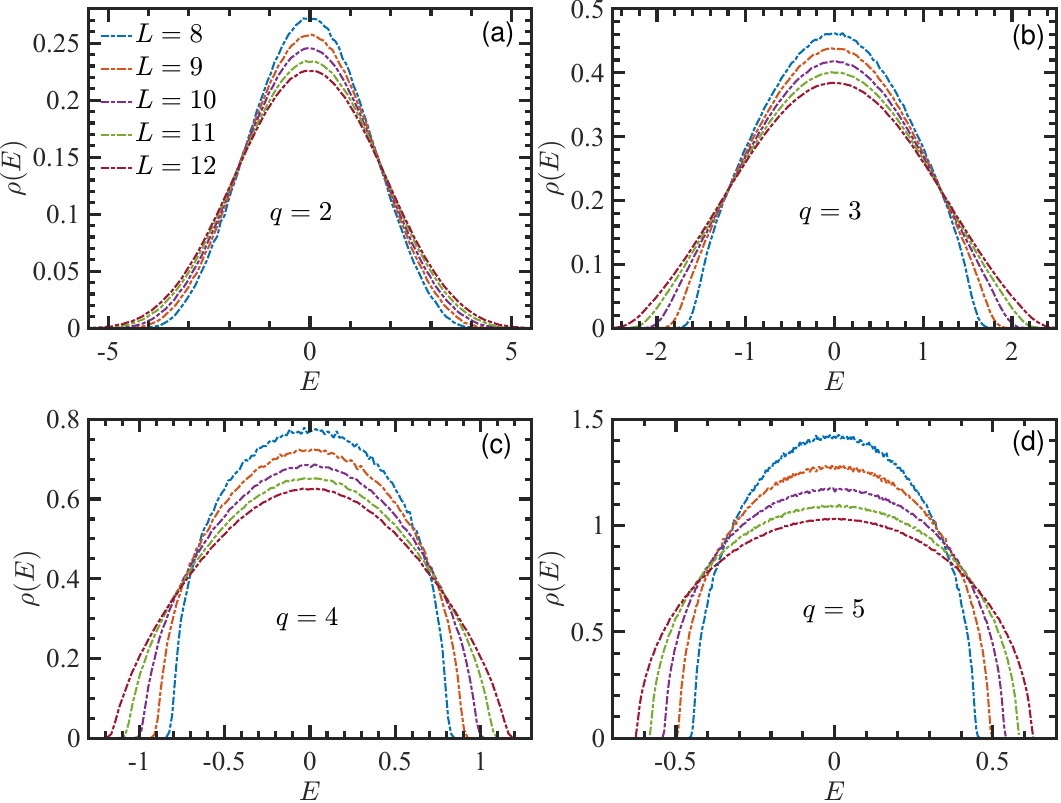}
	\caption{The normalized density of states for spin SYK-$q$ model for different system sizes $L$ with (a) $q=2$, (b) $q=3$, (c) $q=4$, and (d) $q=5$. The Hamiltonian is given by Eq. \eqref{seq:spinsyskham}. For even $q$, the energy eigenvalues are calculated only in the even parity sector, for odd $q$, we calculate energy eigenvalues using full Hilbert space. We use $2^{20-L}$ Hamiltonian realization.}
    \label{sfig:dos_with_prefactor_spinSYK}
\end{figure}

\begin{figure}[ht]
	\centering
	\includegraphics[width=0.7\textwidth]{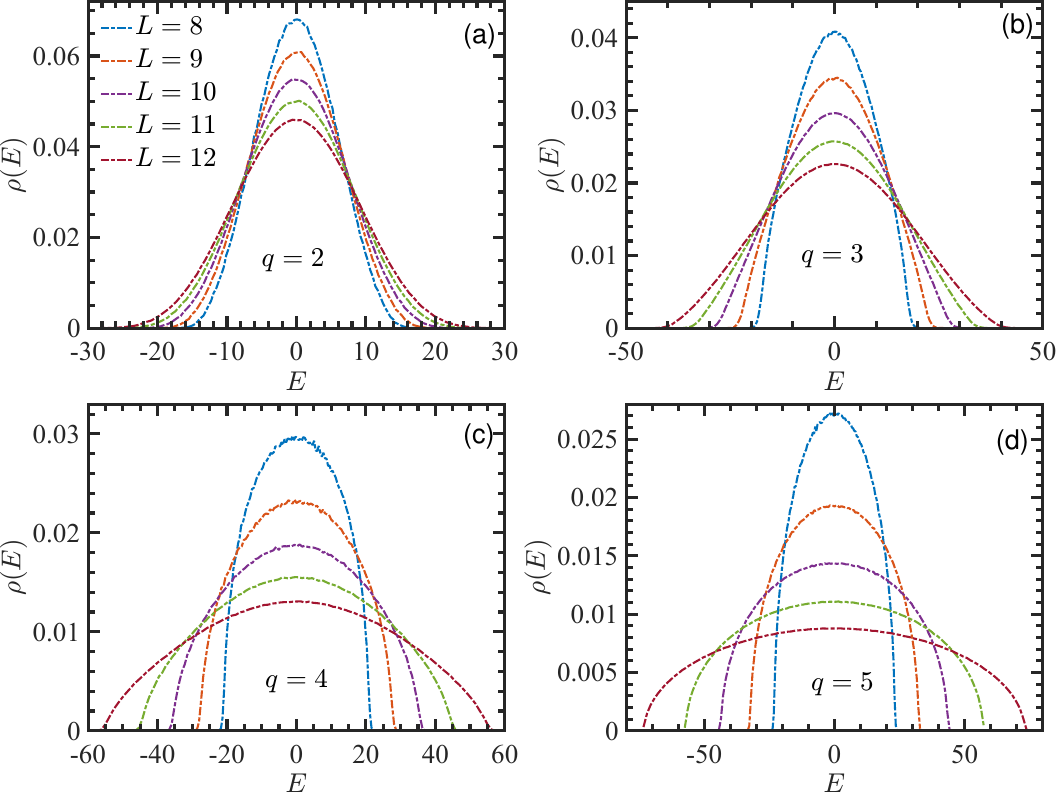}
	\caption{The normalized density of states for spin SYK-$q$ model by setting the prefactor of the Hamiltonian given by Eq. \eqref{seq:spinsyskham} to be \emph{unity} for different system sizes $L$ with (a) $q=2$, (b) $q=3$, (c) $q=4$, and (d) $q=5$. For even $q$, the energy eigenvalues are calculated only in the even parity sector, for odd $q$, we calculate energy eigenvalues using full Hilbert space. We use $2^{20-L}$ Hamiltonian realization.}\label{sfig:dos_with_prefactorunity_spinSYK}
\end{figure}

In this section we report the results of the density of states for spin SYK-$q$ model for $q=2,3,4,5$ and $L=8, \cdots,12$. These results corroborate the findings of the previous results \cite{Hanada_2024} and are presented here for completeness. To gain insight into the global spectral properties of the spin SYK-$q$ model, we examine the normalized density of states (DOS). The DOS characterizes the distribution of many-body energy levels and provides useful information about the spectral structure underlying the dynamics.
Fig.~\ref{sfig:dos_with_prefactor_spinSYK} shows the density of states for the spin SYK-$q$ model with $q=2,3,4$ and $5$ for several system sizes. In Fig.~\ref{sfig:dos_with_prefactorunity_spinSYK} we plot the same but setting the prefactor of the Hamiltonian as unity. 

\begin{figure}[H]
	\centering
	\includegraphics[width=0.7\textwidth]{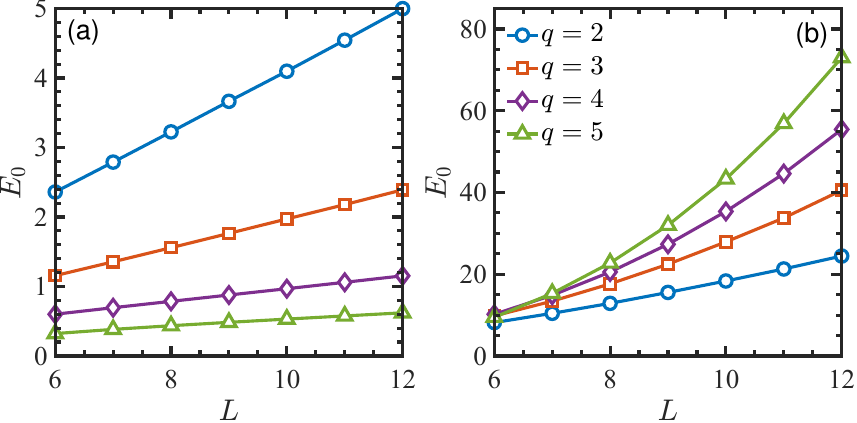}
	\caption{The absolute values of averaged ground-state energy as a function of system size $L$ for different values of the interaction order $q$ for (a) the spin SYK-$q$ model given by Eq. \eqref{seq:spinsyskham} and (b) the spin SYK-$q$ model with the prefactor in Hamiltonian, in Eq. \eqref{seq:spinsyskham} set to unity. For each system size $L$, the results are averaged over $2^{20-L}$ realizations.}\label{sfig:eovsl}
\end{figure}

In Fig \ref{sfig:eovsl} we show the scaling of ground state energy as function of system size $L$ and for various value of $q$. This is then further used to obtain the proper rescaling factor of the Hamiltonian to obtain the energy fixed time scale. To achieve this we use the fact that for ground state energy, $E_{0}$, $\frac{\vert E_{0}\vert}{L}= C(q)$ where $C(q)$ is a $q$ dependent constant. Then implies that for a given value of $q$,  $E_{0}/L$ is a constant line and the extrapolation of that line to the gives us the factor $C(q)$ in the thermodynamic limit. We then rescale our Hamiltonian given by Eq. \eqref{seq:spinsyskham} by $C(q)$.
\clearpage

\clearpage


\bibliography{references}